\documentclass[aps,prx,reprint,superscriptaddress,nofootinbib,longbibliography]{revtex4-2}

\usepackage[utf8]{inputenc}
\usepackage[T1]{fontenc}
\usepackage{lmodern}
\usepackage{microtype}
\usepackage{mathtools}

\usepackage{placeins}
\usepackage{xspace}
\usepackage{braket}
\usepackage{physics}
\usepackage{amsmath}
\usepackage{amsfonts}
\usepackage{graphicx}
\usepackage{hyperref}
\hypersetup{
  colorlinks=true,
  citecolor=blue,
  linkcolor=blue,
  urlcolor=blue
}

\usepackage{mdframed}
\usepackage{comment}
\usepackage{qcircuit}
\usepackage{multirow}
\usepackage{adjustbox}
\usepackage{color}
\usepackage{makecell}

\newcommand{\CNOT}{\textsc{cnot}}
\newcommand{\CZ}{\textsc{cz}}

\providecommand{\CNOT}{\ensuremath{\mathrm{CNOT}}\xspace}

\newcommand{\cgate}[1]{*+<.6em>{#1} \POS ="i","i"+UR;"i"+UL **\dir{-};"i"+DL **\dir{-};"i"+DR **\dir{-};"i"+UR **\dir{-},"i" \cw}
\newcommand{\cmultigate}[2]{*+<1em,.9em>{\hphantom{#2}} \POS [0,0]="i",[0,0].[#1,0]="e",!C *{#2},"e"+UR;"e"+UL **\dir{-};"e"+DL **\dir{-};"e"+DR **\dir{-};"e"+UR **\dir{-},"i" \cw}

\begin{document}

\title{Active Readout Error Mitigation}

\author{Rebecca Hicks$^{\#}$}
\email{rebecca_hicks@berkeley.edu}
\affiliation{Physics Department, University of California, Berkeley, Berkeley, CA 94720, USA}

\author{Bryce Kobrin$^{\#}$}
\email{bkobrin@berkeley.edu}
\affiliation{Physics Department, University of California, Berkeley, Berkeley, CA 94720, USA}
\affiliation{Physics Division, Lawrence Berkeley National Laboratory, Berkeley, CA 94720, USA}

\author{Christian W. Bauer}
\email{cwbauer@lbl.gov}
\affiliation{Physics Division, Lawrence Berkeley National Laboratory, Berkeley, CA 94720, USA}

\author{Benjamin Nachman}
\email{bpnachman@lbl.gov}
\affiliation{Physics Division, Lawrence Berkeley National Laboratory, Berkeley, CA 94720, USA}

\begin{abstract}
Mitigating errors is a significant challenge for near term quantum computers.  One of the most important sources of errors is related to the readout of the quantum state into a classical bit stream.  A variety of techniques have been proposed to mitigate these errors with post-hoc corrections.  We propose a complementary scheme to actively reduce readout errors on a shot-by-shot basis by encoding single qubits,  immediately prior to readout, into multi-qubit states.  The computational resources of our technique are independent of the circuit depth and are compatible with the error rates and connectivity of many current devices.  We analyze the potential of our approach using two types of error-correcting codes and, as a proof of principle, demonstrate an 80\% improvement in readout error on the IBMQ Mumbai quantum computer.
\end{abstract}

\date{\today}
\maketitle

\def\thefootnote{\#}\footnotetext{These authors contributed equally.}

\section{Introduction}

Quantum computers are promising tools to solve many computationally intractable scientific and industrial problems.   However, existing noisy intermediate-scale quantum (NISQ) computers~\cite{Preskill2018quantumcomputingin} suffer from significant errors that must be mitigated before obtaining useful results.  These errors can be categorized as initialization, state preparation, or readout errors.  
Initialization errors, which often arise from thermal noise, lead to residual entropy in the starting state of the system.
State preparation errors result from mis-calibrated quantum gates and unintended couplings with the environment.  
The focus of this paper is readout errors, which result from decoherence during measurement and from the overlapping support between the measured physical quantities that correspond to the $\ket{0}$ and $\ket{1} $ states.  Readout errors can be significant and even dominate the error budget for relatively shallow quantum circuits.

A number of strategies have recently been proposed for mitigating readout errors which rely on classical, post-processing techniques. 
A particularly common scheme involves measuring a response matrix $R_{ij}=\Pr(\text{measure $i$} | \text{true state is $j$})$ and performing regularized matrix inversion on the output distribution~\cite{geller_efficient_2020,song_10-qubit_2017,gong_genuine_2019,wei_verifying_2020,hamilton2020scalable}.
However, this scheme requires characterizing, at worst, an exponentially large matrix, and inverting the response matrix may suffer from numerical instabilities  \cite{geller_efficient_2020}.
%
%
%
%
%
More fundamentally, such passive methods can only reduce the bias from readout errors on average, rather than correcting individual errors on a shot-by-shot basis ~\cite{1910.01969,1904.11935,1907.08518,arute2020quantum,10.1145/3352460.3358265}.
Thus, they cannot be applied to tasks where obtaining individual output states---not expectation values---is desired, including random circuit sampling \cite{arute2019quantum,wu2021strong}, prime factorization \cite{shor1999polynomial}, and high energy physics simulations \cite{nachman2021quantum}. 
%


\begin{figure}[b!]

\[
\Qcircuit @C=1em @R=0.8em {
    \lstick{\ket{0}}  & \gate{U} & \qw &\qw& \meter & \cw \cw[1] &\cgate{R^{-1}} &\cw &\cw \\
    &&&&&\\
    &&&&\big\downarrow&&\\
    &&&&&\\
    \lstick{\ket{0}}  & \gate{U} & \qw &\ctrl{1}& \meter & \cw \cw[1] &\cmultigate{1}{R^{-1}} &\cw &\cw\\
    \lstick{\ket{0}}  & \qw& \qw &\targ &\meter & \cw \cw[1] & \pureghost{R^{-1}} & \cw  & \cw\\
     }
\]
\caption{A schematic diagram of the active readout error mitigation protocol for a single qubit and a quantum circuit represented by the unitary operation $U$. (Top) In passive readout error mitigation, the qubit is read out directly and classical post-processing (box with $R^{-1}$) is applied to reduce the bias in readout errors. (Bottom) In active error correction, the qubit is encoded before readout in a non-local state. The encoding circuit shown implements the 2-qubit repetition code, which enables error detection but not correction. If desired, passive readout error mitigation may be applied in combination with the active strategy, by post-processing the output of the encoded circuit (box with $R^{-1}$).
}
\label{fig:schematic}
\end{figure}
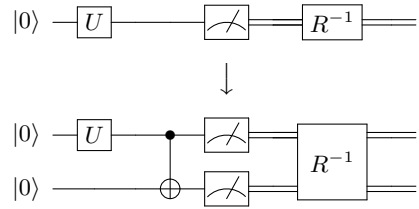

A more powerful strategy for correcting individual errors, during but not limited to readout, is based on the celebrated framework of quantum error correction (QEC). 
The core idea behind QEC is to embed the logical state into non-local degrees of freedom, such that local errors that affect only a few degrees of freedom can be isolated and corrected \cite{Gottesman09anintroduction,terhal2015quantum,Nielsen:2011:QCQ:1972505}. 
To utilize this protection for quantum processing, operations on the logical state must also be encoded and, crucially, be able to withstand a low rate of physical errors; this is the hallmark of \emph{fault-tolerant} quantum computation. 
While individual aspects of error correction / fault tolerance have been demonstrated on small-scale experiments  \cite{errorcorrecting,PhysRevA.97.052313,Barends2014,Kelly2015,Linke:2017, Takita:2017, Roffe:2018, Vuillot:2018, Willsch:2018,Harper:2019,chen2021exponential}, performing full fault-tolerant computation imposes stringent technical requirements which are infeasible for near-term devices \cite{campbell2017roads,gidney2021factor}. 
Moreover, unless all components of a computation are implemented fault-tolerantly, the encoding is likely to have little effect on or even worsen the computational outcome compared to a direct, unencoded implementation.
Finding loopholes to this ``all-or-nothing'' outlook remains an essential task for bridging the gap towards full fault-tolerant computation.

\begin{figure}
    \centering
    \includegraphics[width=0.45\textwidth]{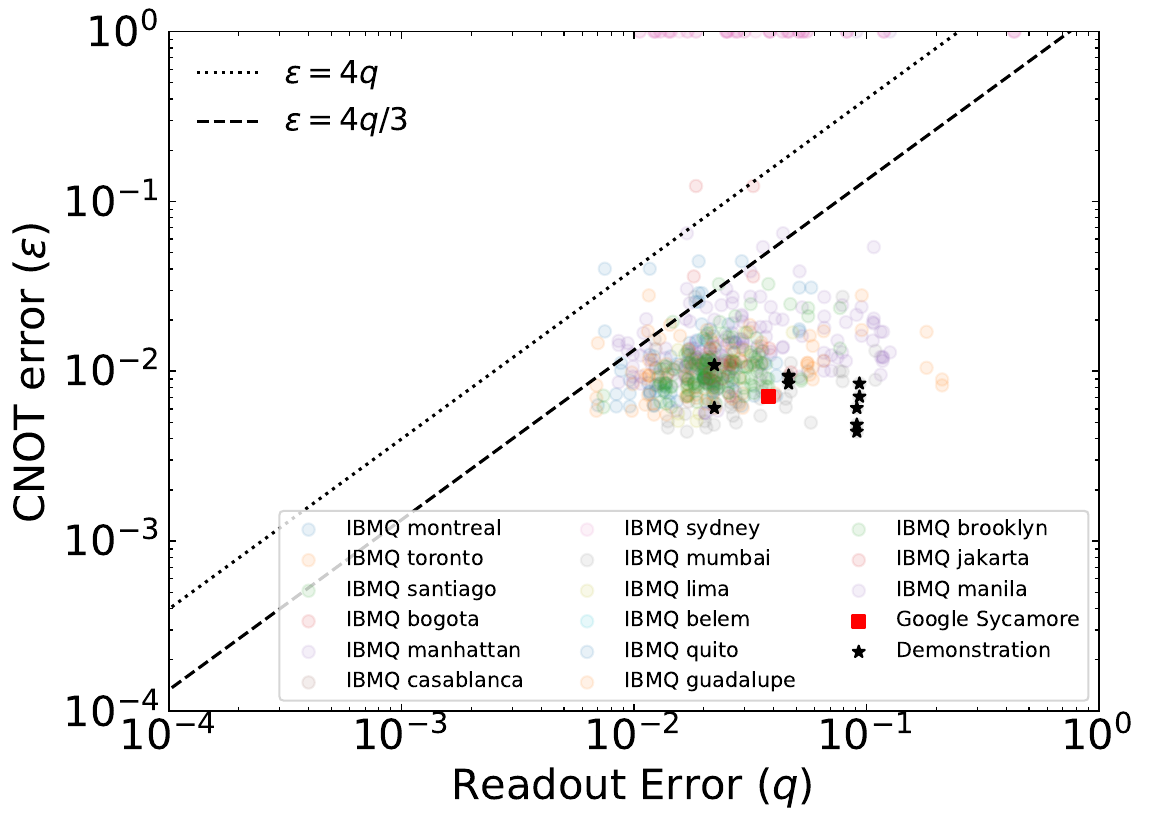}
    \caption{A scatter plot of the \CNOT\ error rates ($\epsilon)$ versus the readout error rates ($q$) across many IBMQ quantum computers.  The dotted line indicates the threshold below which active readout error mitigation is effective when error \emph{detection} is performed, while the dashed line indicates the analogous threshold for error \emph{correction} with the 3-qubit repetition code (see Section \ref{sec:results}). The points labelled `Demonstration' correspond to the five qubits on IBMQ Mumbai used in Sec.~\ref{eq:ibmq}. For the Sycamore device, we consider a decomposition of a \CNOT~gate into a \CZ~gate and two single-qubit Hadamard gates, whose individual error rates are estimated to be $0.41\%$ and $0.15\%$, respectively \cite{Sycamore}.}
    \label{fig:igmqrates}
\end{figure}

In this paper, we propose and analyze an active error mitigation technique that targets readout errors only.
The essence of our approach is to encode each qubit immediately before readout into a multi-qubit logical state, as exemplified in Fig.~\ref{fig:schematic}. 
%
%
%
A general feature of this strategy is a tradeoff between suppressing readout noise and introducing errors during the encoding circuit.
Our strategy is thus most effective on devices whose intrinsic readout errors dominate over entangling gate errors, a situation that arises among many existing quantum computers (Fig.~\ref{fig:igmqrates}).
%
%
Moreover, the resource requirement of the encoding in terms of ancilla qubits and gate overhead is independent of circuit depth, making it well suited for near-term applications.
To this end, we implement our protocol on an IBMQ quantum computer and demonstrate almost an order of magnitude reduction in readout errors.
%
%
%
%

The rest of the paper is organized as follows. In Section~\ref{sec:method}, we introduce the active readout error mitigation protocol after briefly reviewing existing forms of readout error mitigation.  
In Section~\ref{sec:results} and \ref{sec:sim}, we present analytical results and numerical simulations based on simple error models.
In Section \ref{eq:ibmq}, we demonstrate the experimental performance of our strategy on a superconducting quantum computer. 
In Section~\ref{sec:conclusions}, we conclude with a brief summary and outlook on error mitigation strategies.

\section{Method}
\label{sec:method}

Consider a quantum system which is characterized in the computational basis by the frequencies $t_i=\Pr(\text{true state in $i$})= \textrm{Tr}\left[ \rho \dyad i \right ]$, where $i \in \mathbb{Z}^{2^{n_\text{qubits}}}$.
The goal of quantum readout is to accurately and efficiently estimate $t_i$ using a series of measurements on identically prepared quantum systems.
This task becomes non-trivial in the presence of readout errors, since these lead to biases in the measured frequencies $m_i=\Pr(\text{measure $i$})$ compared to the true frequencies $t_i$.

A widely used strategy to correct these biases is to characterize the response matrix $R$ and estimate the true state frequencies as $\hat{t}=R^{-1}m$.
%
 %
As an illustrative example, consider a single qubit quantum state with $t_0 = 1-t_1 = p$ and suppose that it undergoes symmetric readout noise with error probability $q$, i.e. the response matrix is given by $R_{10}=R_{01}=q$ and $R_{00}=R_{11}=1-q$.  
If a series of measurements is performed, the expected frequency in the $0$ state is $\mathbb{E}[m_0] = \lambda$ and its variance is $\text{Var}[m_0] = \lambda(1-\lambda)$ for $\lambda=p+q(1-2p)$.
Performing matrix inversion allows one to correct the average bias.
In particular, applying
\begin{align} \label{eq:passive_corr}
    \hat{t}=\begin{pmatrix}1-q & q \cr q & 1-q\end{pmatrix}^{-1}\begin{pmatrix} m_0\cr m_1\end{pmatrix}\,
\end{align}
leads to the correct expected value of $\hat{t}$, i.e.~$\mathbb{E}[\hat{t}]=(p,1-p)$.

However, this passive approach towards error mitigation is limited in two fundamental ways.
First, the \emph{variance} of $\hat t$ typically remains larger than that of the noise-free case; indeed, in our simple example, the variance increases by an $\mathcal{O}(q)$ factor:
\begin{align}
\begin{split} \label{eq:variance}
  \textrm{Var}[\hat t_0] &= \textrm{Var}[m_0] - 2q \textrm{Cov}[m_0,m_1] + \mathcal{O}(q^2) \\
  &= \lambda(1-\lambda)(1+2q) + \mathcal{O}(q^2) \\
  &= \textrm{Var}[t_0]+q\left[1-2p(1-p)\right] + \mathcal{O}(q^2),
 \end{split}
\end{align}
where we have used that $\textrm{Cov}[m_0,m_1] = -m_0 m_1$ for a multinomial distribution.
Correspondingly, the number measurements required to estimate $\hat t_i$ to a desired precision, which is proportional to $\text{Var}[\hat t_i]$, also increases with the readout error rate.

Second, and more significantly, we have assumed that $m_i$ can be estimated using a large number of measurements relative to the size of the vector space. 
This is generally the case when computing the expectations of quantum observables, where the relevant vector space is that of the observable and not the full Hilbert space. 
But there are many problems of interest where the goal is to sample a relatively sparse number of output states rather, than to estimate statistical properties of the output distributions; for such problems, passive correction methods are fundamentally inapplicable. 

\begin{figure}[b]

\[
\Qcircuit @C=1em @R=0.8em {
    \lstick{\ket{0}}  & \gate{U} & \qw &\qw&\qw& \meter & \cw \cw[1] &\cgate{R^{-1}} &\cw &\cw \\
    &&&&&\\
    &&&&\big\downarrow&&\\
    &&&&&\\
    \lstick{\ket{0}}  & \gate{U} & \qw &\ctrl{1}& \ctrl{2} & \meter & \cw \cw[1] & \cmultigate{2}{R^{-1}} &\cw &\cw\\
    \lstick{\ket{0}}  & \qw& \qw &\targ & \qw&\meter & \cw \cw[1] & \pureghost{R^{-1}} & \cw  & \cw\\
    \lstick{\ket{0}}  & \qw& \qw &\qw &\targ &\meter & \cw \cw[1] & \pureghost{R^{-1}} & \cw  & \cw\\
     }
\]
\caption{An illustration of the repetition (3,1) code, which enables both error detection and error correction. }
\label{fig:schematic2}
\end{figure}
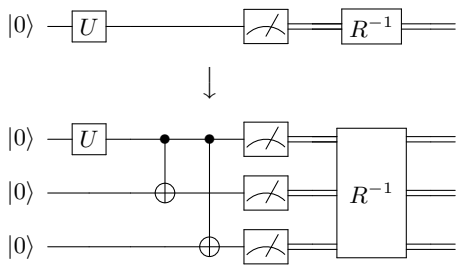

Active readout error mitigation overcomes both of these limitations by effectively reducing the readout error $q$ on a shot-by-shot basis. 
%
%
The main idea is to take a circuit where the physical qubits are the logical qubits and encode the qubits before they are measured into a bigger multiqubit array.  
This encoding is analogous to conventional strategies for quantum error correction, but with two important distinctions. 
First, in the case of readout error mitigation, one is only concerned with bit-flip errors in the computational basis since phase-flip errors do not affect the measured distribution. 
Second, the encoding is performed \emph{after} state preparation rather than at the beginning of the circuit.
These simplifications allow us to circumvent the significant space and gate overhead typically associated with full quantum error correction.

The simplest version of active readout error mitigation is based on the two-qubit repetition code.  
As depicted in Fig.~\ref{fig:schematic}, each qubit is entangled with a unique partner qubit using a single \CNOT~gate (though other fully entangling gates could also be employed).  
Without errors, the measured outcomes are either $00$ or $11$, whereas with a single bit-flip error, the measured outcomes become $01$ or $10$; single qubit readout errors can thus be detected but not corrected. 
%
%
A natural extension of this encoding is to introduce a second ancilla qubit and entangle it with the original qubit (Fig.~\ref{fig:schematic2}). 
This forms a three-qubit repetition code and allows for the correction of single qubit readout errors --- i.e.~by taking the majority vote among the three qubit --- or the detection of two-qubit readout errors.
We henceforth refer to these encodings as the (2,1) and (3,1) codes, where the notation $(n,k)$ indicates that $n$ physical qubits are required to encode $k$ logical qubits.

By design, these encodings offer substantial protection against readout errors; nevertheless, they remain susceptible to certain gate errors that occur during the encoding circuit. 
For example, the single \CNOT~gate in the two-qubit encoding may lead to a correlated bit-flip error on \emph{both} of the qubits, resulting in a spurious measurement outcome despite error detection.
In general, if the average two-qubit error rate is $\epsilon$, one expects an effective readout error rate of $q_\textrm{eff} \approx \alpha \epsilon$, where $\alpha$ is an order-one constant that depends on specific protocol (e.g.~two-qubit vs.~three-qubit, and error detection vs.~correction), as well as the error model for the entangling gates. 
We confirm this scaling for a symmetric depolarizing noise model via analytical results in Sec.~\ref{sec:results} and numerical simulations in Sec.~\ref{sec:sim}.
Active readout error mitigation is thus beneficial whenever the two-qubit error rate is lower than the intrinsic readout error rate $q$. 
In practice, this condition is met by many existing quantum devices, as depicted in Fig.~\ref{fig:igmqrates} for Google Sycamore and a variety of IBMQ quantum computers.

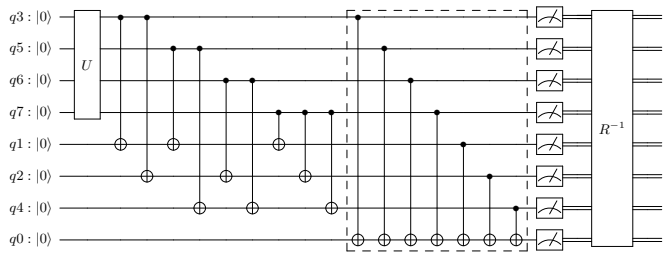
\begin{figure}[t]
\adjustbox{max width=0.45\textwidth}{
\Qcircuit @C=1em @R=0.8em {
    \lstick{q3:\ket{0}}  & \multigate{3}{U}  &\ctrl{4}&\ctrl{5}&\qw&\qw&\qw&\qw&\qw&\qw&\qw& \ctrl{7}&\qw&\qw&\qw&\qw&\qw&\qw&\meter & \cw \cw[1] &\cmultigate{7}{R^{-1}} &\cw &\cw \\
    \lstick{q5:\ket{0}}  & \ghost{U} & \qw & \qw&\ctrl{3}&\ctrl{5}&\qw&\qw&\qw&\qw&\qw&\qw&\ctrl{6}&\qw&\qw&\qw&\qw&\qw& \meter & \cw \cw[1] &\pureghost{R^{-1}} &\cw &\cw \\
    \lstick{q6:\ket{0}}  & \ghost{U}  & \qw& \qw &\qw&\qw&\ctrl{3}&\ctrl{4}&\qw&\qw&\qw&\qw&\qw&\ctrl{5}&\qw&\qw&\qw&\qw& \meter & \cw \cw[1] &\pureghost{R^{-1}} &\cw &\cw \\
    \lstick{q7:\ket{0}}  & \ghost{U} & \qw& \qw &\qw&\qw&\qw&\qw&\ctrl{1}&\ctrl{2}&\ctrl{3}&\qw&\qw&\qw&\ctrl{4}& \qw&\qw&\qw&\meter & \cw \cw[1] &\pureghost{R^{-1}} &\cw &\cw \\
    \lstick{q1:\ket{0}}  & \qw& \targ & \qw&\targ& \qw&\qw&\qw&\targ&\qw&\qw&\qw&\qw&\qw&\qw&\ctrl{3}&\qw&\qw&\meter & \cw \cw[1] &\pureghost{R^{-1}} &\cw &\cw \\
    \lstick{q2:\ket{0}}  & \qw & \qw& \targ &\qw& \qw&\targ&\qw&\qw&\targ&\qw&\qw&\qw&\qw&\qw&\qw&\ctrl{2}&\qw &\meter & \cw \cw[1] &\pureghost{R^{-1}} &\cw &\cw \\
    \lstick{q4:\ket{0}}  & \qw & \qw& \qw &\qw& \targ&\qw &\targ&\qw&\qw&\targ&\qw&\qw&\qw&\qw&\qw&\qw&\ctrl{1}&\meter & \cw \cw[1] &\pureghost{R^{-1}} &\cw &\cw \\
     \lstick{q0:\ket{0}}  & \qw & \qw& \qw &\qw& \qw&\qw &\qw&\qw&\qw&\qw&\targ&\targ&\targ&\targ&\targ&\targ&\targ&\meter & \cw \cw[1] &\pureghost{R^{-1}} &\cw &\cw \gategroup{1}{12}{8}{18}{.7em}{--} \\
      }
}
\caption{Illustration of the encoding for the Hamming (8,4) code. The first four qubits (indexed q3, q5, q6, q7) contain logical state information while the remaining qubits are the parity bits. For the Hamming (7,4) code we omit the last parity bit (q0) and all gates connected to it, i.e.~those within the black dashed box. }
\label{fig:schematic3}
\end{figure}

Generalizing our strategy, one may consider encoding circuits for implementing arbitrary \emph{classical} error correction codes.
To do so, one would add ancilla qubits and entangle them with the original qubits to generate a classical code in the computational basis, i.e.~each bitstring on the original qubits is mapped to an encoded bitstring on the full set of qubits. 
To understand the tradeoffs of using increasingly complex codes, we compare two families of error encoding schemes: the aforementioned repetition codes and two versions of the Hamming code---the (7,4) and (8,4) codes---illustrated in Fig.~\ref{fig:schematic3}.  
%
%
We test the performance of these codes via numerical simulations in Sec.~\ref{sec:sim} and summarize their key differences in Table \ref{tab:Tab1}.
In particular, we find that both types of codes offer comparable levels of error mitigation, and the more important factor for determining the effective error rate is whether error detection or error correction is performed.
%

\begin{table}
\begin{tabular}{c|c|c|c|c|c}
  \hline \hline
  \multirow{2}{*}{Encoding} & \multirow{2}{*}{$(n,k)$ } & \multicolumn{1}{c|}{Det.~or} & \multicolumn{2}{c|}{Eff.~error rate} & \multicolumn{1}{c}{Discarded} \\  & & cor.? & $q$-dependence & $\alpha$ & measurements? \\ \hline
  \makecell{Repetition\\code}  & \makecell{(2,1)\\(3,1)\\(3,1)} & \makecell{det.\\det.\\cor.} & \makecell{$\sim q^2$\\$\sim q^3$\\$\sim q^2$} & \makecell{$1/4$\\$1/4$\\$3/4$} &  \makecell{Yes\\Yes\\No} \\ \hline
  \makecell{Hamming\\code}  & \makecell{(7,4)\\(7,4)\\(8,4)\\(8,4)} & \makecell{det.\\cor.\\det.\\hybrid} & \makecell{$\sim q^3$\\$\sim q^2$\\$\sim q^4$\\$\sim q^3$} & \makecell{$1/4$\\$7/8$\\$1/4$\\$1/4$} & \makecell{Yes\\No\\Yes\\Yes} \\ 
  \hline \hline
\end{tabular}
\caption{Summary of the encoding schemes presented in this work. The notation $(n,k)$ indicates that $n$ physical qubits are required to encode $k$ logical qubits. The effective error rate $Q_{\textrm{eff}}$ [defined in Eq.~\eqref{eq:Q_eff}] scales non-linearly with the nominal readout error rate $q$, and is linearly proportional to the \CNOT~error rate $\epsilon$ with a susceptibility $\alpha$ given by $Q_{\textrm{eff}} / k \approx \alpha \epsilon$. }
\label{tab:Tab1}
\end{table}

\section{Analytical Results}
\label{sec:results}

\subsection{Two-qubit repetition code}
\label{sec:two-qubit}

We begin by analyzing the simplest realization of active readout error mitigation: protecting a single qubit via the two-qubit encoding depicted in Fig.~\ref{fig:schematic}.  
Since each qubit is encoded separately, this represents an excellent model of a full quantum circuit, ignoring cross talk.
Suppose that after applying $U$, the quantum state (for the first qubit) is
\begin{align} \label{eq:psi}
  \rho = \dyad{\psi}, \quad \ket \psi = \sqrt{p} \ket 0 + \sqrt{1-p} \ket 1\,,
\end{align}
where, without loss of generality, we have chosen the phase to be real and positive.
Moreover, as in the previous section, suppose that readout errors are described by a symmetric bit-flip channel with $\Pr(1 \to 0)=q$ and $\Pr(0 \to 1)=q$ for all qubits. 
Finally, we model the dominant error channel affecting the \CNOT~gate as a two-qubit depolarizing noise channel:
\begin{align} \label{eq:cnot}
  \rho\mapsto (1-\epsilon)\CNOT\,\rho\,\CNOT + \frac{\epsilon}{4}I_4\,,
\end{align}
where $I_4$ is the identity matrix. 
Note that this last assumption, while widely used in theoretical analysis, is not essential to our qualitative results \cite{error_model}. 

Based on this setup, we can compare the rate of readout errors with and without the encoding. 
%
%
In the nominal case with no encoding, the measurement frequencies are given by
\begin{align} \label{eq:nominal}
\begin{split}
  m_0  &= p + (1-2p) q \\
  m_1  &= (1-p) - (1-2p)q
 \end{split}
\end{align}
where $m_i = \textrm{Tr}\left[ \rho \dyad i \right ]$.
For the encoded circuit (bottom of Fig.~\ref{fig:schematic}), we instead have four output states with frequencies
\begin{align}
\begin{split}
  m_{00} &\approx p (1-2q-\epsilon)+q^2+ \frac \epsilon 4\\
  m_{01} &\approx q - q^2 + \frac \epsilon 4 \\
  m_{10} &\approx q - q^2 + \frac \epsilon 4 \\
  m_{11} &\approx (1-p) (1-2q-\epsilon)+q^2+ \frac \epsilon 4 ,
 \end{split}
\end{align}
where we have neglected $\mathcal{O}(\epsilon q)$ terms and higher.
In post-selection, we keep only the two symmetric states, $\ket{\bar 0} = \ket{00}$ and $\ket{\bar 1} = \ket{11}$, whose normalized frequencies are
\begin{align}
\begin{split}
\label{eq:two_qubit_Q}
  m_{\bar 0} &= \frac{m_{00}}{m_{00}+m_{11}} \approx p+\left(\frac{\epsilon}4+q^2\right) (1-2p) \\
  m_{\bar 1} &= \frac{m_{11}}{m_{00}+m_{11}} \approx (1-p)-\left(\frac{\epsilon}4+q^2\right) (1-2p).
 \end{split}
\end{align}
Comparing to the nominal case (Eq.~\ref{eq:nominal}), the final distribution is characterized by an effective error rate, $q_\textrm{eff} \approx \epsilon/4+q^2$, which is independent of $p$.
Thus, for $q^2 \ll \epsilon$, the encoding effectively lowers the readout error rate by a factor of $\sim 4 q/\epsilon$.  

Physically, this leading order dependence on $\epsilon$ may be understood through a simple error analysis.
From a stochastic perspective, the depolarizing channel corresponds to introducing a random two-qubit Pauli error, each with probability $\epsilon/16$.
There are 4 errors which could lead to a undetected measurement error, i.e.~\textsc{XX, XY, YX}, and \textsc{YY}; thus, the total effective error rate is $\epsilon/4$.
Additional mechanisms that result in an undetected error correspond to ($i$) two independent measurement errors or ($ii$) a \CNOT\ error followed by a measurement error. 
However, these are subleading with probability $\mathcal{O}(q^2)$ and $\mathcal{O}(q\epsilon)$, respectively. 

 
A few additional remarks are in order.
First, as discussed in the previous section, it is important to consider not only the average bias in the measurment outcome but also the number of repetitions required to achieve a desired precision.
For the unencoded circuit, the required repetitions is directly proportional to the variance of the measurement outcome distribution (Eq.~\ref{eq:variance}).
However, in the two-qubit encoding, a fraction of the repetitions are discarded (i.e.~those in the $01$ and $10$ states), thereby contributing to the total repetitions but not to the output distribution. 
Thus, there is a competition between the improved bias offered by the encoding and the repetition overhead of the discarded states; whether or not this competition favors the encoding depends quantitatively on the desired observable, error model, and initial state. 

Second, it is instructive to consider the case of non-symmetric readout errors.
If we directly generalize our scheme to a non-symmetric error channel, $\Pr(1 \to 0) = q(1+\kappa)$ and $\Pr(0 \to 1) = q(1-\kappa)$, we find that the normalized probabilities, $m_{\bar 0}$ and $m_{\bar 1}$, exhibit a systematic bias compared to the true probabilities at linear order in $q|\kappa|$.
%
Notably, this bias does not result from measurement errors that flip between $\ket {00} \leftrightarrow \ket{11}$, but rather from the fact that one of these states may be more likely to decay into the non-symmetric sector and therefore discarded during post-selection \cite{asymm_noise}.
 
An \emph{ad hoc} solution to eliminate this bias to perform error detection as before but, in a random half of the repetitions, apply an extra \textsc{X} gate immediately before the measurement. 
In doing so, the resulting distribution is \emph{rebalanced}, or rendered symmetric with respect to the initial state \cite{Hicks_2021}. 
The system can then be modeled as having an effective measurement rate, $\Pr(1 \to 0) = \Pr(0 \to 1) = 1/2[\Pr(1 \to 0) + \Pr(0 \to 1)] = q$, regardless of the asymmetry parameter $\alpha$.
Of course, this scheme assumes that the X gate itself introduces negligible errors compared to readout errors or two-qubit errors, which is reasonable assumption for many experimental platforms.

As we discuss next, an alternative approach to mitigate the effects of asymmetric noise---as well as eliminate the discarded measurements---is to perform error \emph{correction} on the encoding qubit.
%
%


%

\subsection{Three-qubit repetition code}
\label{sec:two-qubit}

By adding a third qubit to the encoded circuit, as shown in Fig.~\ref{fig:schematic2}, one can realize the repetition (3,1) code whose output states without errors are $000$ and $111$.
With this code, two bit-flip errors can be detected, while single bit-flip errors can be corrected by taking the ``majority vote'' among the three qubits. 
To understand the tradeoffs between error detection and detection, we analyze the three-qubit encoding circuit under the previous error model, i.e.~symmetric readout errors ($\Pr(1 \to 0) = \Pr(0 \to 1) = q$) and two-qubit depolarizing errors (at rate $\epsilon$) following each CNOT gate.
 
We begin by computing the final measurement distribution up to order $\mathcal{O}(q e)$ and $\mathcal{O}(q^3)$:
\begin{align}
\begin{split}
  m_{000} &\approx \left(1-3q+3q^2-\frac {7\epsilon} 4 \right)p+ \frac \epsilon 4  \\
  m_{001} &\approx  \frac \epsilon 4 p + pq+(1-3p)q^2 \\
  m_{010} &\approx \frac \epsilon 4 (2-p) + pq +(1-3p)q^2\\
  m_{011} &\approx  \frac \epsilon 4 (1-p) + (1-p)q-(2-3p)q^2 \\
  m_{100} &\approx \frac \epsilon 4 p + pq+(1-3p)q^2 \\
  m_{101} &\approx \frac \epsilon 4 (1+p) + (1-p)q -(2-3p)q^2\\
  m_{110} &\approx \frac \epsilon 4 (1-p) + (1-p)q -(2-3p)q^2\\
  m_{111} &\approx \left(1-3q+3q^2-\frac {7\epsilon} 4 \right)(1-p)+ \frac \epsilon 4 
  \end{split}
\end{align}
For error detection, we proceed in analogy with the two-qubit code by computing the normalized frequencies in the logical code subspace:
\begin{align}
\begin{split}
\label{eq:three_qubit_Q}
  m^{\textrm{det}}_{\bar 0} &= \frac{m_{000}}{m_{000}+m_{111}} \approx p+\frac{\epsilon}4(1-2p)\\
  m^{\textrm{det}}_{\bar 1} &= \frac{m_{111}}{m_{000}+m_{111}} \approx (1-p)-\frac \epsilon 4 (1-2p).
 \end{split}
\end{align}
The effective error rate is thus $q^{\textrm{det}}_\textrm{eff} = \epsilon/4 + \mathcal{O}(q^3)$ and exhibits the same leading order dependence on $\epsilon$ as the two-qubit code.
This implies that the three-qubit encoding is subject to failure under the same number of individual gate errors, as we confirm by direct examination of the encoding circuit (see Appendix \ref{app:gate}). 
In contrast, the $q^2$ terms have vanished since three independent readout errors are required to produce an undetectable logical error.

We next simulate error correction by assigning the subspace of states with two or more 0's as $\ket {\bar 0}$, and the remaining states as $\ket {\bar 1}$.
This results in a final output distribution
\begin{align} \label{eq:3-qubit correction}
\begin{split}
  m^{\textrm{cor}}_{\bar 0} &= m_{000}+m_{001}+m_{010}+m_{100} \\
  &\approx p+3\left(\frac{\epsilon}4 +q^2\right) (1-2p) \\
  m^{\textrm{cor}}_{\bar 1} &= m_{111}+m_{110}+m_{101}+m_{110} \\
  &\approx (1-p)-3\left(\frac{\epsilon}4 +q^2\right)(1-2p),
\end{split}
\end{align}
characterized by an effective error rate, $q^{\textrm{cor}}_\textrm{eff} \approx 3(\epsilon/4+q^2)$.
Notably, the dependence on $\epsilon$ is enhanced by a factor of 3 compared to error detection compared with either the 2-qubit or 3-qubit code. 
This difference arises from the increased number of gate errors that can lead to a falsely \emph{corrected} output state.
For example, a single bit-flip on the first qubit after the first \CNOT~gate can \emph{propagate} across the second \CNOT~gate into a two-qubit error, thereby corrupting the measurement outcome (see Appendix \ref{app:gate}).  

Despite the less favorable effective error rate, we emphasize that error correction offers two main advantages compared to error detection, owing to the fact that none of the measurements are discarded during post-processing. 
First, the total number of events kept is larger, leading to reduced statistical uncertainties (see Appendix \ref{app:var}).
More importantly, contrary to error detection schemes, error correction is intrinsically robust against asymmetries in the readout error channels. 
%
%
For example, if the first qubit was subject to an asymmetric readout channel, $\Pr(1 \to 0) = q(1+\kappa)$ and $\Pr(0 \to 1) = q(1-\kappa)$ with $\kappa > 0$, then population would be transferred from $\ket {000} \rightarrow \ket{100}$ more frequently than from $\ket {111} \rightarrow \ket{011}$, leading to a bias in error detection schemes.
But neither of these processes affect the sums, $m_{\bar 0}$ and $m_{\bar 1}$, at leading order in $q$, implying that this bias is eliminated in error correction.

\begin{figure*}[t]
\centering
\includegraphics[width=0.6\textwidth]{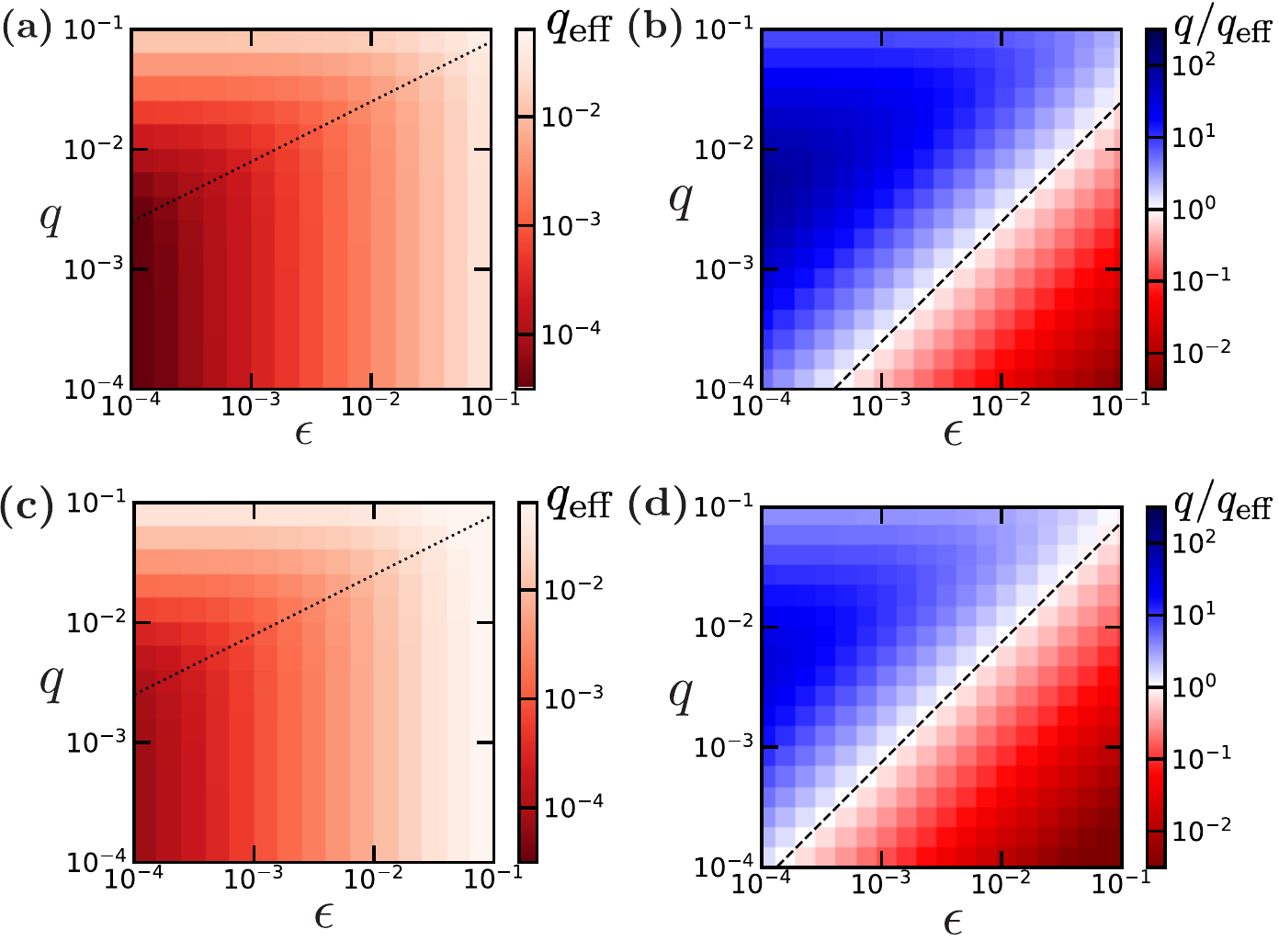}
\caption{
Effective readout error, $q_\textrm{eff}$, as a function of the bare readout error rate, $q$, and the two-qubit error rate, $\epsilon$. (a,b) Readout mitigation is performed via error detection with the two-qubit repetition code. (c,d) Analogous results for error correction with the three-qubit repetition code. In the left column, $q_\textrm{eff}$ is plotted directly and exhibits a weak dependence on $q$ for $\epsilon \gtrsim 16 q^2$ (dotted line). In the right column, $q_\textrm{eff}$ is normalized by the bare readout error rate, $q$. The dashed lines indicate the predicted thresholds for which error mitigation is advantageous, i.e.~(b) $\epsilon = 4q$ for error detection, and (d) $\epsilon = \frac 4 3 q$ for error correction. The results for $q_\textrm{eff}$ are independent of the initial state parameter, $p$.}
\label{fig:avg_error}
\end{figure*}

\section{Simulations} \label{sec:sim}

\subsection{Repetition codes}

To demonstrate active error mitigation with realistic error rates, we perform numerical simulations on the encoded circuits shown in Fig.~\ref{fig:schematic} and Fig.~\ref{fig:schematic2}. 
As before, we consider two types of errors: ($i$) two-qubit depolarizing errors with rate $\epsilon$ applied after each \CNOT~ gate (Eq.~\ref{eq:cnot}), and ($ii$) symmetric readout errors on each qubit with rate $q$.
We implement the noisy circuits using the software package \texttt{Cirq}~\cite{cirq} and measure the outcome distribution directly as $m_i = \textrm{Tr}\left[ \rho \dyad i \right ]$. 
 
For the two-qubit code, we perform error \emph{detection} by discarding the probability of measuring the qubits in the odd parity subspace.
For the three-qubit code, we perform error \emph{correction} on the outcome distribution via majority vote (see Eq.~\ref{eq:3-qubit correction}).
Subsequently, we characterize the effective error rate $q_\textrm{eff}$ using
\begin{align} \label{eq:Q_eff}
\begin{split}
  m_{\bar 0}  &= p + (1-2p) q_\textrm{eff} \\
  m_{\bar 1}  &= (1-p) - (1-2p)q_\textrm{eff}
\end{split}
\end{align}
where $m_{\bar 0/ \bar 1}$ are the error-mitigated probabilities, and $p$ is the initial state parameter defined in Eq.~\ref{eq:psi}.

In Fig.~\ref{fig:avg_error}, we depict the effective error rate for the two encodings as a function of $\epsilon$ and $q$. 
In both cases, for $\epsilon \gg q^2$, $q_\textrm{eff}$ scales linearly with $\epsilon$ and has a negliglible dependence on $q$. 
This is consistent with our leading order analysis and indicates that higher order effects arise primarily from two-qubit readout errors with probability $\mathcal{O}(q^2)$.
For a quantitative comparison with the unencoded readout error rate, we next plot the ratio $q_{\textrm{eff}}/q$ in Fig.~\ref{fig:avg_error}(b).
For low error rates, this ratio crosses unity at $\epsilon=4q$ for error detection and $\epsilon=4q/3$ for error correction, in ageement with our predictions; small deviations from these trends become noticeable as $\epsilon, q \rightarrow 1$.
%
%

Note that although the simulations were performed with a specific value of $p$, the effective error rate $q_\textrm{eff}$ is in fact independent of the initial state.
This is a consequence of using \emph{symmetric} noise channels for the readout and gate error and would change for a state-dependent noise model, e.g.~for readout errors where $\Pr(1 \to 0) \neq \Pr(0 \to 1)$.
 
\subsection{Comparison with the Hamming code}


We now compare the performance of the repetition codes to a more complex encoding circuit based on the Hamming (7,4) code, which encodes 4 logical qubits into 7 physical qubits (Fig.~\ref{fig:schematic3}).
The Hamming code is thus more space-efficient encoding for 4 logical qubits than either repetition code; in fact, even larger Hamming codes reduce the ratio of physical to logical qubits further, quickly approaching an optimal ratio of one (see Appendix \ref{app:Hamming}).
Note, however, that implementing a Hamming code generally requires each logical qubit to be entangled with several ancillary qubits, which requires \textsc{Swap} gates on systems with limited connectivity.

Like the three-qubit repetition code, the 7-qubit Hamming code enables the correction of single-qubit errors and the detection of two-qubit errors. 
This similarity arises because both codes have a \emph{code distance} of three, which refers to the minimum number of local bit-flips (or Hamming distance) that connects two logical states. 
Indeed, it is well known that a classical code with code distance $d$ enables the detection of up to $d-1$ independent errors and the correction of up to $d/2$ errors.
%

A circuit for implementing the Hamming (7,4) code is shown in Fig.~\ref{fig:schematic3}. 
The first four qubits contain the logical state, while the remaining three qubits are ancilla qubits required for the encoding.
Without errors, the state of each ancilla qubit is equal to the parity of a certain set of logical qubits.
In contrast, if a single bit-flip error occurs (on either the logical bits or the ancilla bits), one or more of these conditions will be violated.
A convenient way to check for errors is by multiplying the output state $s$ by the parity check matrix:
\begin{align} \label{eq:parity check}
H = 
\begin{pmatrix}
1 & 1 & 0 & 1 & 1 & 0 & 0 \\
1 & 0 & 1 & 1 & 0 & 1 & 0 \\
0 & 1 & 1 & 1 & 0 & 0 & 1 \\
\end{pmatrix}_{[7,4]},
\end{align}
where the ordering matches the physical layout of the first seven qubits in Fig.~\ref{fig:schematic3}.
Each element for which the bitwise product $Hs$ is equal to 1 (modulo 2) indicates that an error has occurred on either the parity bit or the associated set of logical qubits.
Based on these checks, one can uniquely determine the location of a \emph{single} bit-flip error or detect the occurence of up to two bit-flip errors (see Appendix \ref{app:Hamming}).

\begin{figure}
    \centering
    \includegraphics[width=0.48\textwidth]{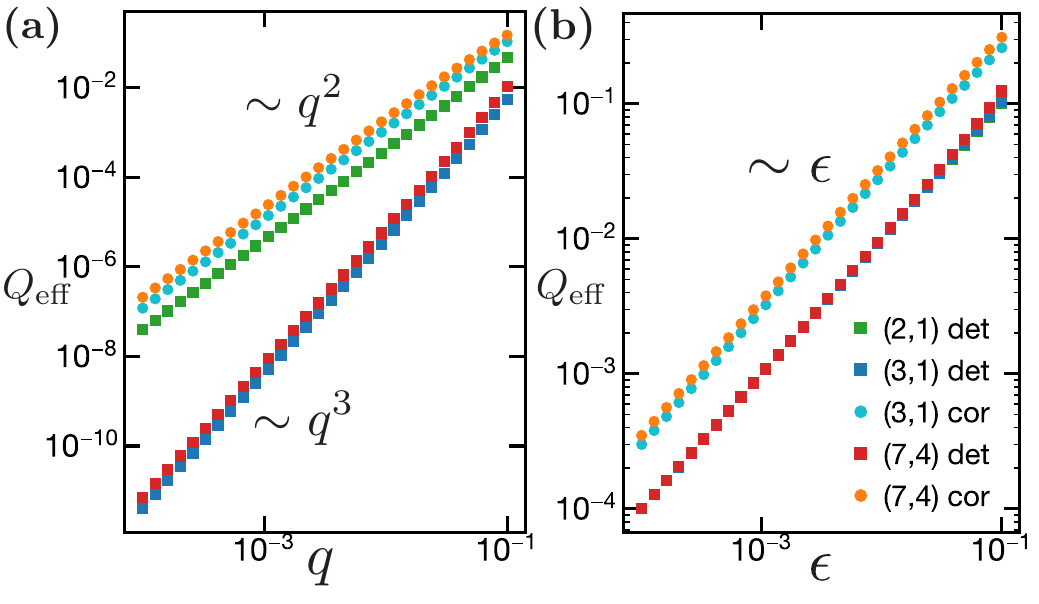}
    \caption{Total readout error rate, $Q_{\textrm{eff}}$, for various error mitigation strategies. (a) $Q_{\textrm{eff}}$ exhibits a power-law scaling as a function the bare readout error rate $q$, with an exponent determined by the code distance and whether error correction or error detection is performed. (b) $Q_{\textrm{eff}}$ depends linearly on the \CNOT~gate error rate, $\epsilon$. The linear prefactors $\alpha$, defined by $Q_{\textrm{eff}}/k = \alpha \epsilon$, are summarized in Table \ref{tab:Tab1}.}
    \label{fig:err_scaling}
\end{figure}

To compare the performance of the Hamming (7,4) code to the repetition codes, we perform noisy simulations based each encoding and apply error either error correction or detection to the output distribution. 
%
%
We estimate the total error rate after error mitigation by computing 
\begin{equation}\label{eq:Q_eff}
    Q_{\textrm{eff}} = 1- \frac{\sum_i R_{ii}}{2^k},
\end{equation} where $R_{ii}$ are the diagonal elements of the response matrix, i.e.~the probabilities of measuring the correct state.
This generalizes our previous error rate $q_\textrm{eff}$ (Eq.~\ref{eq:Q_eff}); indeed, $Q_{\textrm{eff}} = q_\textrm{eff}$ in the case of a single qubit.

We begin by simulating the encoded circuits with readout errors only, i.e.~taking $\epsilon = 0$. 
As shown in Fig.~\ref{fig:err_scaling}(a), the total error rate for both the Hamming (7,4) code and the repetition (3,1) code scales as $\sim q^2$ when using error correction and $\sim q^3$ when using error detection. 
These trends follow directly from having a code distance of $d=3$, which implies that 2 or 3 independent bit-flip errors are required for a logical errors in the two respective cases. 
Similarly, the total error rate for the two-qubit repetition code scales as $\sim q^2$, since its code distance is $d=2$. 

We next isolate the effect of gate errors by plotting the error rate as a function of $\epsilon$ [Fig.
\ref{fig:err_scaling}(b)].
In contrast to the dependence on readout errors, all encoding schemes lead to the same general linear scaling, i.e.~$Q_{\textrm{eff}} \approx 4 \alpha \epsilon$ (we include a factor of 4 for the number of logical qubits). 
This indicates that a \emph{single} gate error during the encoding can propagate into logical errors on the measurement outcomes, as we had shown previously for the repetition codes.
More specifically, for all cases of error detection, we observe the same prefactor $\alpha = 1/4$; whereas, for error correction, $\alpha$ is an order-one factor larger.
In fact, we previously derived $\alpha = 3/4$ for a single qubit encoded with the (3,1) code (see Eq.~\eqref{eq:3-qubit correction}). 
Similarly, we can determine $\alpha = 7/8$ for the Hamming (7,4) code by counting the relevant error channels (see Appendix \ref{app:gate}).
%

\begin{figure}
    \centering
    \includegraphics[width=0.4\textwidth]{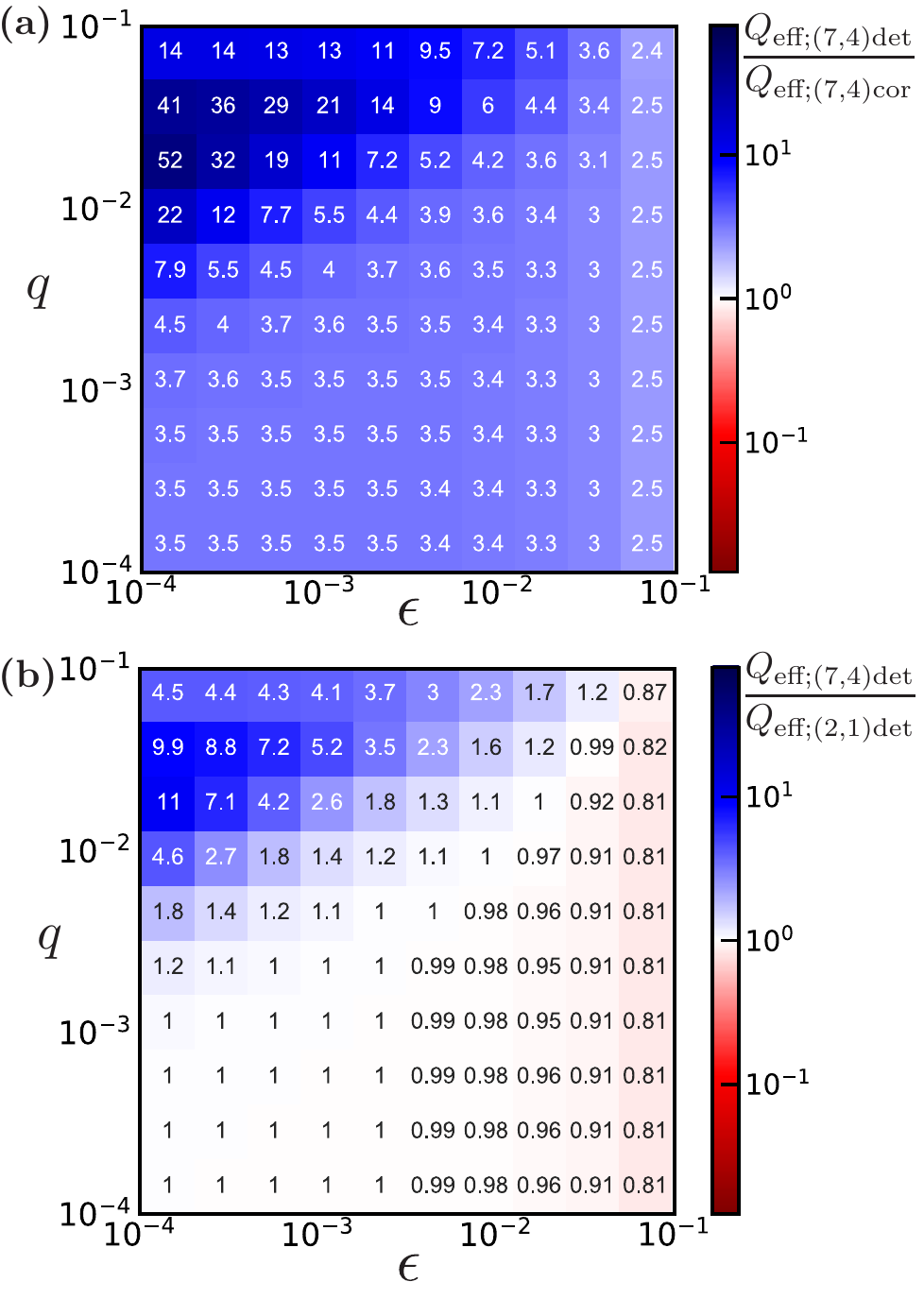}
    \caption{Comparison of the effective error rate $Q_{\textrm{eff}}$ between various active error mitigation strategies. (a) Ratio of $Q_\textrm{eff}$ of the (7,4) code with error detection to that of the (7,4) code with error correction. (b) Analogous ratio of the (7,4) code with error detection to the (2,1) code with error detection.}
    \label{fig:code-comparison}
\end{figure}

We now probe the effects of both readout and gate errors through two direct comparisons.
First, we compare the 7-qubit encoding circuit using both error detection and error correction [Fig.~\ref{fig:code-comparison}(a)].
For the most part, their total error rates differ by a factor of $\sim 3$ due to their respective scalings with $\epsilon$; this ratio increases further when $\epsilon \ll q$, owing to the favorable scaling with $q$ offered by error detection.
%
%
Second, we compare error detection using the 7-qubit encoding and the 2-qubit encoding [Fig.~\ref{fig:code-comparison}(a)]. 
Analagous with the previous case, the error rates are nearly equivalent for the two encodings except when $\epsilon \ll q$, in which case the scaling with $q$ becomes relevant.



Finally, in Appendix \ref{app:8-hamming}, we analyze an extended (8,4) version of the Hamming code which includes an extra ancilla qubit compared to the (7,4) code and has an increased code distance of $d=4$.
Interestingly, there are two natural circuits that lead to the same classical encoding and differ in the number of entangling gates. 
While one would naively expect the circuit with the fewest gates to have the lowest error rate, we find the opposite to be true. 
This highlights the importance of designing encoding circuits that are robust not only to readout errors, but also to gate errors that occur during the encoding circuit. 


\section{Experimental demonstration}
\label{eq:ibmq}

\begin{figure}
    \centering
    \includegraphics[width=0.5\textwidth]{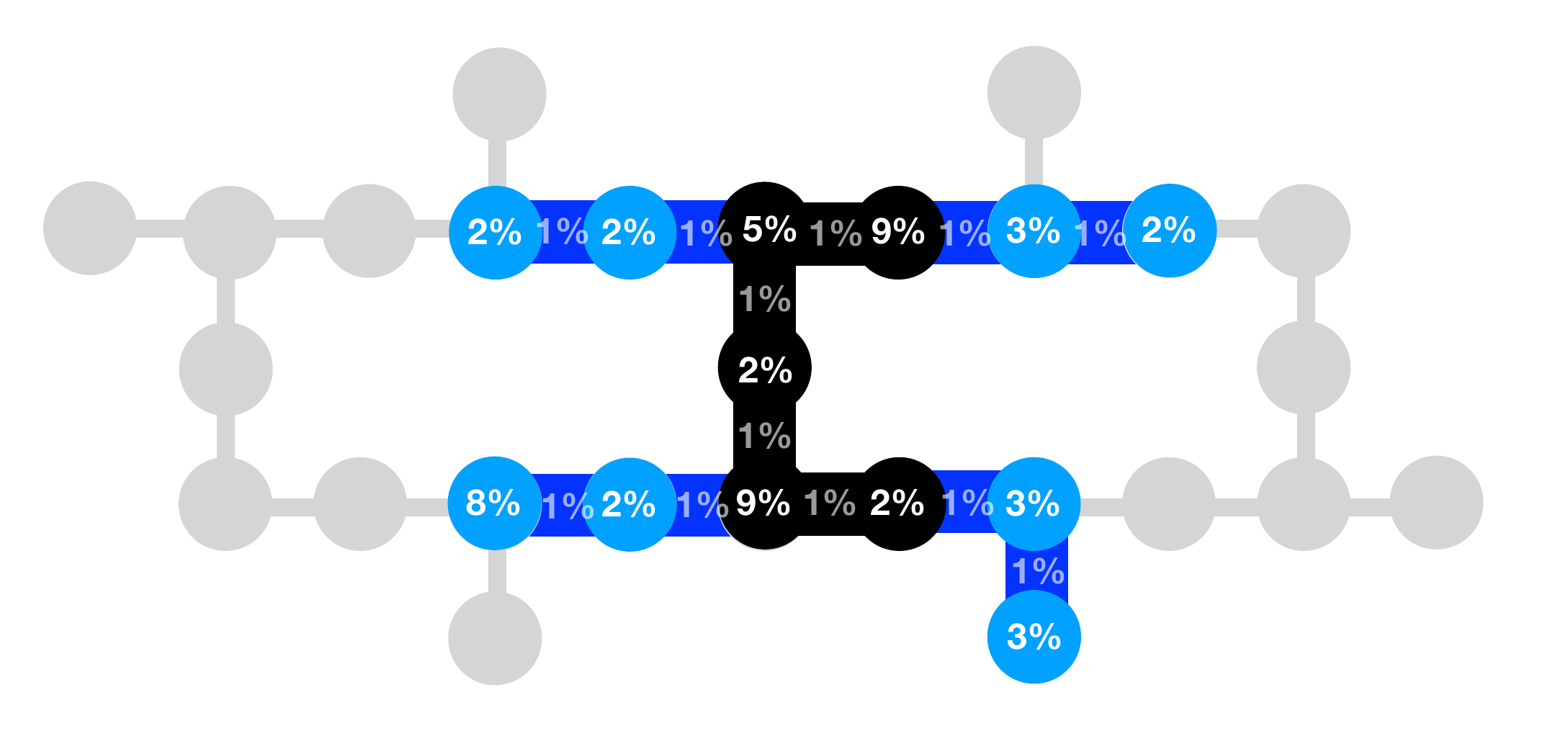}
    \caption{A diagram of the IBMQ Mumbai computer layout, where circles represent qubits and links represent qubits that are connected.  The qubits used for the measurement presented in Fig.~\ref{fig:igmqresults} are colored in black (logical qubits) and blue (coding qubits).  The readout errors are reported in the circles and the two qubit errors are reported on links connecting the circles.  Error rates are reported from August 19, 2021.}
    \label{fig:ibmqmanhattan}
\end{figure}

This section demonstrates the experimental performance of the repetition code on the IBMQ Mumbai quantum computer.  This computer has 27 qubits total, arranged in a pattern depicted in Fig.~\ref{fig:ibmqmanhattan}.  To demonstrate the active readout error correction protocol, we construct a 5 qubit sub-computer consisting of the five filled black circles in Fig.~\ref{fig:ibmqmanhattan} (corresponding to qubits 12-16 in the computer's labeling scheme).  Due to the adjacency map of connected qubits, we are unable to encode all qubits, without adding extra \textsc{Swap} gates.  Instead, the first (top right filled black circle in Fig.~\ref{fig:ibmqmanhattan}), second, fourth and fifth (counter-clockwise from the first) are encoded with the (3,1) repetition code to improve the readout errors. 

\begin{figure}
    \centering
    \includegraphics[width=0.45\textwidth]{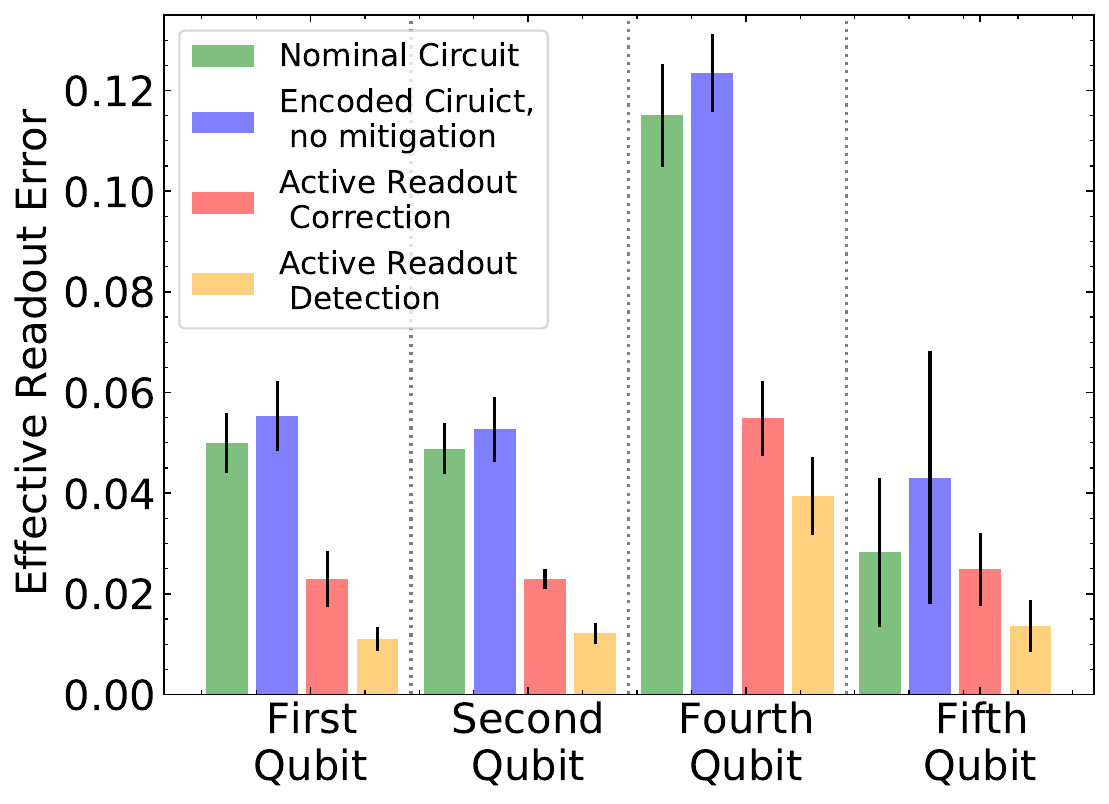}
    \caption{The effective readout error for the first (top right filled black circle in Fig.~\ref{fig:ibmqmanhattan}), second, fourth and fifth (counter-clockwise from the first) are encoded with the (3,1) repetition code to improve the readout errors. For each qubit, the left (green) bar corresponds to the nominal circuit without any additional \CNOT{s}, the blue represents the circuit with the additional qubits and \CNOT{s} without using them for correction, the red shows the results with the active readout error correction, and the orange shows the results with active readout error detection.  The bar height is the average over all $2^5=32$ initial states and the error bar represents the standard deviation.}
    \label{fig:effectiverate}
\end{figure}

In Fig.~\ref{fig:effectiverate}, the effective readout error rates for the four qubits are compared under three different scenarios.
First, we measure the readout error rate with the encoding circuit but \emph{without} performing error mitigation (i.e.~we discard the measurements of the ancilla bits).
As expected, this leads to an increase in the readout error rate relative to that of the nominal circuit.
Indeed, the observed increase of $\sim 1\%$ is consistent with the independently measured rate of depolarizing errors (Fig.~\ref{fig:ibmqmanhattan}).  
%

Second, we measure the effective error rate after performing either error detection or correction.
With either scheme, we observe a substantial improvement in the error rate, e.g.~dropping by a factor of five in the first two qubits compared to the unencoded qubits.
This indicates that the suppression of readout errors due to the encoding outweighs the errors introduced by the entangling gates and is consistent with the relatively large readout error rates for these qubits (Fig.~\ref{fig:ibmqmanhattan}).
%
%

%



\begin{figure*}
    \centering
    \includegraphics[width=0.75\textwidth]{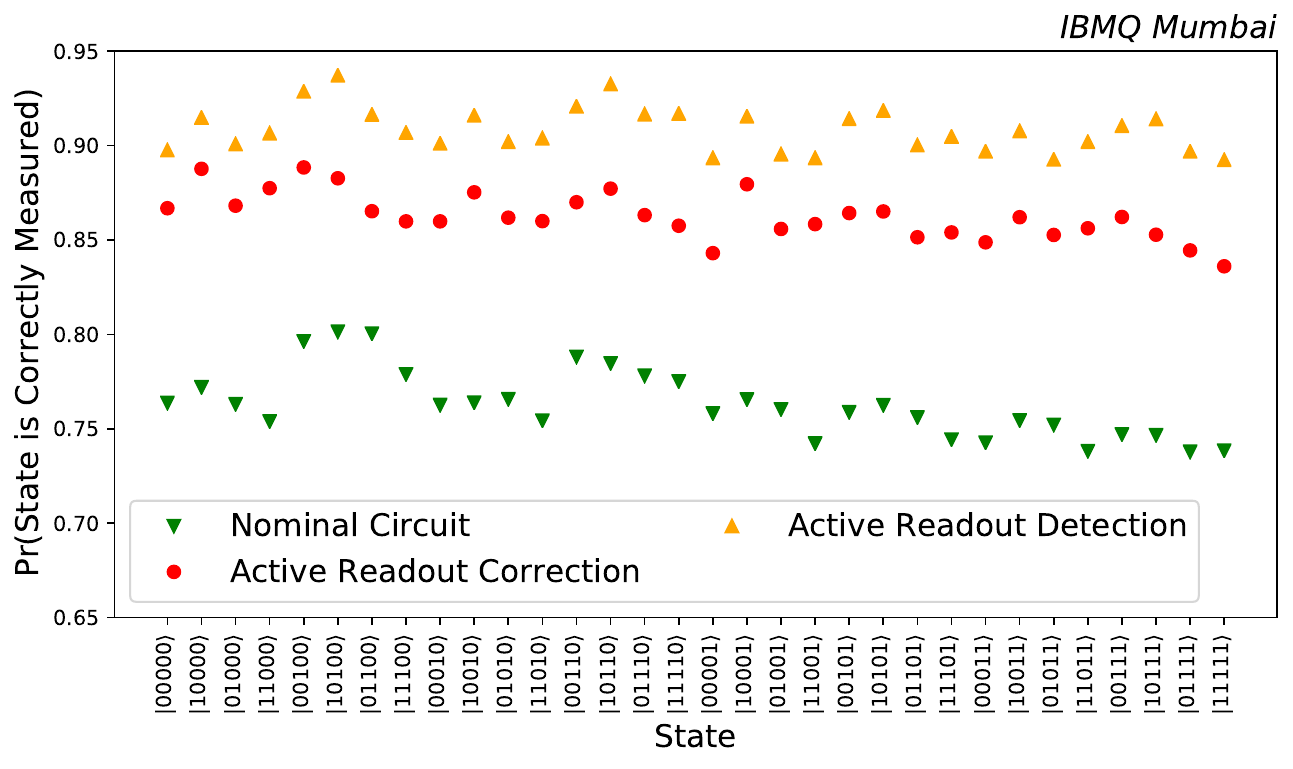}
    \caption{The diagonal elements of the response matrix measured on IBMQ Manhattan for an effective five logical qubit setup (see Fig.~\ref{fig:ibmqmanhattan}) with and without active readout mitigation.  Due to the layout of IBMQ Manhattan, only two of the qubits have readout error correction and one has readout error detection.  The other two qubits are not encoded.}
    \label{fig:igmqresults}
\end{figure*}

A global picture of the subcomputer performance is illustrated in Fig.~\ref{fig:igmqresults}.  Even though only four of the five qubits are encoded, the probability for a prepared state to be correctly measured increases from $\sim$75\% to more than 90\% on average.  

On other quantum computers, the number and location of qubits that may be encoded for readout mitigation will be determined by the device connectivity. 
This is a relatively minor concern for problems where only a small number of qubits are read out at a time (e.g.~when measuring few-body correlation functions), and thus only a few additional qubits are required for the readout encoding. 
These ancilla qubits may be spare nearby qubits on the device or, for greater space efficiency, they may be qubits that are initially involved in the computation and repurposed directly before readout, i.e.~by resetting to the ground state.  
 
A more challenging situation occurs when measuring \emph{all} the qubits simultaneously is required, e.g.~as in random circuit sampling. 
Indeed, with current two-dimensional grid architectures, applying our scheme to all qubits is only well suited to computational tasks performed on a one-dimensional (or quasi-one-dimensional) subset of qubits, such that a neighboring row (or few rows) of qubits may be used for the readout encoding.
Looking forward, we envision that improvements in hardware connectivity will overcome this limitation. 
For example, one may consider designing a two-dimensional device with two sublattices of qubits: one sublattice containing the computational qubits and the other sublattice containing ancilla qubits. 
The computational qubits would be connected to each other as in current devices but also to a single ancilla qubit in the opposite sublattice for implementing the (2,1) code. 
Notably, as the ancilla qubits would only involved in the readout operations, they would not affect the implementation or performance of the main computational task (assuming negligible cross-talk).

\section{Conclusions}
In this work, we proposed a scheme for active readout error mitigation based on encoding the output state of a quantum circuit via a classical error correcting code. 
We showed that this approach generally provides significant readout improvement on devices whose bare readout error rate is comparable or larger than the error rate of entangling gates.

More specifically, we introduced two forms of encoding (the repetition code and the Hamming code) and analyzed the tradeoffs between error detection and error correction.
%
%
%
Which scheme gives the optimal results depends on the criteria that are most important to satisfy.
Error correction schemes are favorable if a large asymmetry in the readout error exists and it can not be mitigated through rebalancing techniques. 
On the other hand, error detection schemes have a smaller overall effective readout error.
Among the error detection schemes, one should choose a specific encoding depending relative size of readout and \CNOT\ gate errors, as well as how many ancillary qubits are available. 
If readout errors dominate over gate errors, then correction schemes with a code distance of 3, namely the (7,4) Hamming code and the (3,1) repetition code are preferable. 
On the other hand, if gate errors dominate, then all detection codes are more or less equal.
Finally, the (7,4) Hamming code requires the least amount of ancillary qubits for a given number of physical qubits (however the encoding circuit involves non-local entangling gates).
%

With any of these implementations, active readout mitigation offers a few general advantages compared to passive error mitigation techniques.
First, active error mitigation does not require characterizing a device's noise parameters (e.g.~response matrix), which can involve an exponential overhead and be subject to temporal drifts; nor does it suffer from pathologies that can occur when applying post-processing (e.g.~matrix inversion) on specific error models. 
Second, by removing errors on a shot-by-shot basis, our approach is more applicable to tasks where sampling individual states is desired and, for general tasks, can be more effective at reducing not only the average measurement bias but also its variance. 
Crucially, active and passive strategies can also be combined to maximize the readout fidelity; i.e.~by actively encoding the readout qubits and implementing error detection / correction, followed by post-processing the effective output distribution using passive correction techniques.

Lastly, in contrast to active \emph{gate} error correction schemes, active readout error correction is fully compatible with near-term quantum hardware.
Indeed, the protocol we have presented in this paper only requires local encoding whereby additional qubits need only have a small number of connections with entangling gates. 
Active readout error mitigation thus provides a practical intermediate step on the long path towards realizing full quantum error correction.

\emph{Note added:} After this work had been completed, we became aware of a recent study \cite{gunther2021improving} by G\"unther et.~al.~which also introduces active readout correction.  Aside from the core concept, the two papers are complementary in their analysis and proposed implementations. 



\label{sec:conclusions}

\section*{Code and Data}

The code and data for this paper can be found at \url{https://github.com/LBNL-HEP-QIS/activereadouterrors}.

 \begin{acknowledgments}
 We would like to thank Marat Freytsis, Maurice Garcia-Sciveres, Maxwell Block, Francisco Machado and Jarrod McClean for useful discussions and feedback on the manuscript.  We thank Vince Pascuzzi for his IBMQ device data script.  This work is supported by the U.S. Department of Energy, Office of Science under contract DE-AC02-05CH11231. In particular, support comes from Quantum Information Science Enabled Discovery (QuantISED) for High Energy Physics (KA2401032) and the Office of Advanced Scientific Computing Research (ASCR) through the Accelerated Research for Quantum Computing Program.   This research used resources of the Oak Ridge Leadership Computing Facility, which is a DOE Office of Science User Facility supported under Contract DE-AC05-00OR22725. 
\end{acknowledgments}

\bibliography{myrefs}

\begin{thebibliography}{37}%
\makeatletter
\providecommand \@ifxundefined [1]{%
 \@ifx{#1\undefined}
}%
\providecommand \@ifnum [1]{%
 \ifnum #1\expandafter \@firstoftwo
 \else \expandafter \@secondoftwo
 \fi
}%
\providecommand \@ifx [1]{%
 \ifx #1\expandafter \@firstoftwo
 \else \expandafter \@secondoftwo
 \fi
}%
\providecommand \natexlab [1]{#1}%
\providecommand \enquote  [1]{``#1''}%
\providecommand \bibnamefont  [1]{#1}%
\providecommand \bibfnamefont [1]{#1}%
\providecommand \citenamefont [1]{#1}%
\providecommand \href@noop [0]{\@secondoftwo}%
\providecommand \href [0]{\begingroup \@sanitize@url \@href}%
\providecommand \@href[1]{\@@startlink{#1}\@@href}%
\providecommand \@@href[1]{\endgroup#1\@@endlink}%
\providecommand \@sanitize@url [0]{\catcode `\\12\catcode `\$12\catcode
  `\&12\catcode `\#12\catcode `\^12\catcode `\_12\catcode `\%12\relax}%
\providecommand \@@startlink[1]{}%
\providecommand \@@endlink[0]{}%
\providecommand \url  [0]{\begingroup\@sanitize@url \@url }%
\providecommand \@url [1]{\endgroup\@href {#1}{\urlprefix }}%
\providecommand \urlprefix  [0]{URL }%
\providecommand \Eprint [0]{\href }%
\providecommand \doibase [0]{https://doi.org/}%
\providecommand \selectlanguage [0]{\@gobble}%
\providecommand \bibinfo  [0]{\@secondoftwo}%
\providecommand \bibfield  [0]{\@secondoftwo}%
\providecommand \translation [1]{[#1]}%
\providecommand \BibitemOpen [0]{}%
\providecommand \bibitemStop [0]{}%
\providecommand \bibitemNoStop [0]{.\EOS\space}%
\providecommand \EOS [0]{\spacefactor3000\relax}%
\providecommand \BibitemShut  [1]{\csname bibitem#1\endcsname}%
\let\auto@bib@innerbib\@empty
\bibitem [{\citenamefont {Preskill}(2018)}]{Preskill2018quantumcomputingin}%
  \BibitemOpen
  \bibfield  {author} {\bibinfo {author} {\bibfnamefont {J.}~\bibnamefont
  {Preskill}},\ }\bibfield  {title} {\bibinfo {title} {Quantum {C}omputing in
  the {NISQ} era and beyond},\ }\href
  {https://doi.org/10.22331/q-2018-08-06-79} {\bibfield  {journal} {\bibinfo
  {journal} {{Quantum}}\ }\textbf {\bibinfo {volume} {2}},\ \bibinfo {pages}
  {79} (\bibinfo {year} {2018})}\BibitemShut {NoStop}%
\bibitem [{\citenamefont {Geller}\ and\ \citenamefont
  {Sun}(2020)}]{geller_efficient_2020}%
  \BibitemOpen
  \bibfield  {author} {\bibinfo {author} {\bibfnamefont {M.~R.}\ \bibnamefont
  {Geller}}\ and\ \bibinfo {author} {\bibfnamefont {M.}~\bibnamefont {Sun}},\
  }\bibfield  {title} {\bibinfo {title} {Efficient correction of multiqubit
  measurement errors},\ }\href {http://arxiv.org/abs/2001.09980} {\bibfield
  {journal} {\bibinfo  {journal} {arXiv:2001.09980 [quant-ph]}\ } (\bibinfo
  {year} {2020})},\ \bibinfo {note} {arXiv: 2001.09980}\BibitemShut {NoStop}%
\bibitem [{\citenamefont {Song}\ \emph {et~al.}(2017)\citenamefont {Song},
  \citenamefont {Xu}, \citenamefont {Liu}, \citenamefont {Yang}, \citenamefont
  {Zheng}, \citenamefont {Deng}, \citenamefont {Xie}, \citenamefont {Huang},
  \citenamefont {Guo}, \citenamefont {Zhang} \emph
  {et~al.}}]{song_10-qubit_2017}%
  \BibitemOpen
  \bibfield  {author} {\bibinfo {author} {\bibfnamefont {C.}~\bibnamefont
  {Song}}, \bibinfo {author} {\bibfnamefont {K.}~\bibnamefont {Xu}}, \bibinfo
  {author} {\bibfnamefont {W.}~\bibnamefont {Liu}}, \bibinfo {author}
  {\bibfnamefont {C.-p.}\ \bibnamefont {Yang}}, \bibinfo {author}
  {\bibfnamefont {S.-B.}\ \bibnamefont {Zheng}}, \bibinfo {author}
  {\bibfnamefont {H.}~\bibnamefont {Deng}}, \bibinfo {author} {\bibfnamefont
  {Q.}~\bibnamefont {Xie}}, \bibinfo {author} {\bibfnamefont {K.}~\bibnamefont
  {Huang}}, \bibinfo {author} {\bibfnamefont {Q.}~\bibnamefont {Guo}}, \bibinfo
  {author} {\bibfnamefont {L.}~\bibnamefont {Zhang}}, \emph {et~al.},\
  }\bibfield  {title} {\bibinfo {title} {10-qubit entanglement and parallel
  logic operations with a superconducting circuit},\ }\href@noop {} {\bibfield
  {journal} {\bibinfo  {journal} {Physical review letters}\ }\textbf {\bibinfo
  {volume} {119}},\ \bibinfo {pages} {180511} (\bibinfo {year}
  {2017})}\BibitemShut {NoStop}%
\bibitem [{\citenamefont {Gong}\ \emph {et~al.}(2019)\citenamefont {Gong},
  \citenamefont {Chen}, \citenamefont {Zheng}, \citenamefont {Wang},
  \citenamefont {Zha}, \citenamefont {Deng}, \citenamefont {Yan}, \citenamefont
  {Rong}, \citenamefont {Wu}, \citenamefont {Li}, \citenamefont {Chen},
  \citenamefont {Zhao}, \citenamefont {Liang}, \citenamefont {Lin},
  \citenamefont {Xu}, \citenamefont {Guo}, \citenamefont {Sun}, \citenamefont
  {Castellano}, \citenamefont {Wang}, \citenamefont {Peng}, \citenamefont {Lu},
  \citenamefont {Zhu},\ and\ \citenamefont {Pan}}]{gong_genuine_2019}%
  \BibitemOpen
  \bibfield  {author} {\bibinfo {author} {\bibfnamefont {M.}~\bibnamefont
  {Gong}}, \bibinfo {author} {\bibfnamefont {M.-C.}\ \bibnamefont {Chen}},
  \bibinfo {author} {\bibfnamefont {Y.}~\bibnamefont {Zheng}}, \bibinfo
  {author} {\bibfnamefont {S.}~\bibnamefont {Wang}}, \bibinfo {author}
  {\bibfnamefont {C.}~\bibnamefont {Zha}}, \bibinfo {author} {\bibfnamefont
  {H.}~\bibnamefont {Deng}}, \bibinfo {author} {\bibfnamefont {Z.}~\bibnamefont
  {Yan}}, \bibinfo {author} {\bibfnamefont {H.}~\bibnamefont {Rong}}, \bibinfo
  {author} {\bibfnamefont {Y.}~\bibnamefont {Wu}}, \bibinfo {author}
  {\bibfnamefont {S.}~\bibnamefont {Li}}, \bibinfo {author} {\bibfnamefont
  {F.}~\bibnamefont {Chen}}, \bibinfo {author} {\bibfnamefont {Y.}~\bibnamefont
  {Zhao}}, \bibinfo {author} {\bibfnamefont {F.}~\bibnamefont {Liang}},
  \bibinfo {author} {\bibfnamefont {J.}~\bibnamefont {Lin}}, \bibinfo {author}
  {\bibfnamefont {Y.}~\bibnamefont {Xu}}, \bibinfo {author} {\bibfnamefont
  {C.}~\bibnamefont {Guo}}, \bibinfo {author} {\bibfnamefont {L.}~\bibnamefont
  {Sun}}, \bibinfo {author} {\bibfnamefont {A.~D.}\ \bibnamefont {Castellano}},
  \bibinfo {author} {\bibfnamefont {H.}~\bibnamefont {Wang}}, \bibinfo {author}
  {\bibfnamefont {C.}~\bibnamefont {Peng}}, \bibinfo {author} {\bibfnamefont
  {C.-Y.}\ \bibnamefont {Lu}}, \bibinfo {author} {\bibfnamefont
  {X.}~\bibnamefont {Zhu}},\ and\ \bibinfo {author} {\bibfnamefont {J.-W.}\
  \bibnamefont {Pan}},\ }\bibfield  {title} {\bibinfo {title} {Genuine 12-qubit
  entanglement on a superconducting quantum processor},\ }\href
  {https://doi.org/10.1103/PhysRevLett.122.110501} {\bibfield  {journal}
  {\bibinfo  {journal} {Physical Review Letters}\ }\textbf {\bibinfo {volume}
  {122}},\ \bibinfo {pages} {110501} (\bibinfo {year} {2019})},\ \bibinfo
  {note} {arXiv: 1811.02292}\BibitemShut {NoStop}%
\bibitem [{\citenamefont {Wei}\ \emph {et~al.}(2020)\citenamefont {Wei},
  \citenamefont {Lauer}, \citenamefont {Srinivasan}, \citenamefont
  {Sundaresan}, \citenamefont {McClure}, \citenamefont {Toyli}, \citenamefont
  {McKay}, \citenamefont {Gambetta},\ and\ \citenamefont
  {Sheldon}}]{wei_verifying_2020}%
  \BibitemOpen
  \bibfield  {author} {\bibinfo {author} {\bibfnamefont {K.~X.}\ \bibnamefont
  {Wei}}, \bibinfo {author} {\bibfnamefont {I.}~\bibnamefont {Lauer}}, \bibinfo
  {author} {\bibfnamefont {S.}~\bibnamefont {Srinivasan}}, \bibinfo {author}
  {\bibfnamefont {N.}~\bibnamefont {Sundaresan}}, \bibinfo {author}
  {\bibfnamefont {D.~T.}\ \bibnamefont {McClure}}, \bibinfo {author}
  {\bibfnamefont {D.}~\bibnamefont {Toyli}}, \bibinfo {author} {\bibfnamefont
  {D.~C.}\ \bibnamefont {McKay}}, \bibinfo {author} {\bibfnamefont {J.~M.}\
  \bibnamefont {Gambetta}},\ and\ \bibinfo {author} {\bibfnamefont
  {S.}~\bibnamefont {Sheldon}},\ }\bibfield  {title} {\bibinfo {title}
  {Verifying {Multipartite} {Entangled} {GHZ} {States} via {Multiple} {Quantum}
  {Coherences}},\ }\href {https://doi.org/10.1103/PhysRevA.101.032343}
  {\bibfield  {journal} {\bibinfo  {journal} {Physical Review A}\ }\textbf
  {\bibinfo {volume} {101}},\ \bibinfo {pages} {032343} (\bibinfo {year}
  {2020})},\ \bibinfo {note} {arXiv: 1905.05720}\BibitemShut {NoStop}%
\bibitem [{\citenamefont {Hamilton}\ \emph {et~al.}(2020)\citenamefont
  {Hamilton}, \citenamefont {Kharazi}, \citenamefont {Morris}, \citenamefont
  {McCaskey}, \citenamefont {Bennink},\ and\ \citenamefont
  {Pooser}}]{hamilton2020scalable}%
  \BibitemOpen
  \bibfield  {author} {\bibinfo {author} {\bibfnamefont {K.~E.}\ \bibnamefont
  {Hamilton}}, \bibinfo {author} {\bibfnamefont {T.}~\bibnamefont {Kharazi}},
  \bibinfo {author} {\bibfnamefont {T.}~\bibnamefont {Morris}}, \bibinfo
  {author} {\bibfnamefont {A.~J.}\ \bibnamefont {McCaskey}}, \bibinfo {author}
  {\bibfnamefont {R.~S.}\ \bibnamefont {Bennink}},\ and\ \bibinfo {author}
  {\bibfnamefont {R.~C.}\ \bibnamefont {Pooser}},\ }\href@noop {} {\bibinfo
  {title} {Scalable quantum processor noise characterization}} (\bibinfo {year}
  {2020}),\ \Eprint {https://arxiv.org/abs/2006.01805} {arXiv:2006.01805
  [quant-ph]} \BibitemShut {NoStop}%
\bibitem [{\citenamefont {B.~Nachman}(2020)}]{1910.01969}%
  \BibitemOpen
  \bibfield  {author} {\bibinfo {author} {\bibfnamefont {W.~d. J. C.~B.}\
  \bibnamefont {B.~Nachman}, \bibfnamefont {M.~Urbanek}},\ }\bibfield  {title}
  {\bibinfo {title} {{Unfolding Quantum Computer Readout Noise.}},\ }\bibfield
  {journal} {\bibinfo  {journal} {npj Quantum Information}\ }\textbf {\bibinfo
  {volume} {6}},\ \href {https://doi.org/10.1038/s41534-020-00309-7}
  {10.1038/s41534-020-00309-7} (\bibinfo {year} {2020}),\ \Eprint
  {https://arxiv.org/abs/1910.01969} {arXiv:1910.01969 [hep-ph]} \BibitemShut
  {NoStop}%
\bibitem [{\citenamefont {Yanzhu~Chen}\ and\ \citenamefont
  {Wei}(2019)}]{1904.11935}%
  \BibitemOpen
  \bibfield  {author} {\bibinfo {author} {\bibfnamefont {S.~Y.}\ \bibnamefont
  {Yanzhu~Chen}, \bibfnamefont {Maziar~Farahzad}}\ and\ \bibinfo {author}
  {\bibfnamefont {T.-C.}\ \bibnamefont {Wei}},\ }\href@noop {} {\bibinfo
  {title} {Detector tomography on ibm 5-qubit quantum computers and mitigation
  of imperfect measurement}},\ \bibinfo {howpublished} {arXiv} (\bibinfo {year}
  {2019}),\ \Eprint {https://arxiv.org/abs/1904.11935} {arXiv:1904.11935
  [quant-ph]} \BibitemShut {NoStop}%
\bibitem [{\citenamefont {Filip B.~Maciejewski}\ and\ \citenamefont
  {Oszmaniec}(2019)}]{1907.08518}%
  \BibitemOpen
  \bibfield  {author} {\bibinfo {author} {\bibfnamefont {Z.~Z.}\ \bibnamefont
  {Filip B.~Maciejewski}}\ and\ \bibinfo {author} {\bibfnamefont
  {M.}~\bibnamefont {Oszmaniec}},\ }\href@noop {} {\bibinfo {title} {Mitigation
  of readout noise in near-term quantum devices by classical post-processing
  based on detector tomography}},\ \bibinfo {howpublished} {arXiv} (\bibinfo
  {year} {2019}),\ \Eprint {https://arxiv.org/abs/1907.08518} {arXiv:1907.08518
  [quant-ph]} \BibitemShut {NoStop}%
\bibitem [{\citenamefont {Arute}\ \emph {et~al.}(2020)\citenamefont {Arute}
  \emph {et~al.}}]{arute2020quantum}%
  \BibitemOpen
  \bibfield  {author} {\bibinfo {author} {\bibfnamefont {F.}~\bibnamefont
  {Arute}} \emph {et~al.},\ }\href@noop {} {\bibinfo {title} {Quantum
  approximate optimization of non-planar graph problems on a planar
  superconducting processor}} (\bibinfo {year} {2020}),\ \Eprint
  {https://arxiv.org/abs/2004.04197} {arXiv:2004.04197 [quant-ph]} \BibitemShut
  {NoStop}%
\bibitem [{\citenamefont {Tannu}\ and\ \citenamefont
  {Qureshi}(2019)}]{10.1145/3352460.3358265}%
  \BibitemOpen
  \bibfield  {author} {\bibinfo {author} {\bibfnamefont {S.~S.}\ \bibnamefont
  {Tannu}}\ and\ \bibinfo {author} {\bibfnamefont {M.~K.}\ \bibnamefont
  {Qureshi}},\ }\bibfield  {title} {\bibinfo {title} {Mitigating measurement
  errors in quantum computers by exploiting state-dependent bias},\ }in\ \href
  {https://doi.org/10.1145/3352460.3358265} {\emph {\bibinfo {booktitle}
  {Proceedings of the 52nd Annual IEEE/ACM International Symposium on
  Microarchitecture}}},\ \bibinfo {series and number} {MICRO ’52}\ (\bibinfo
  {publisher} {Association for Computing Machinery},\ \bibinfo {address} {New
  York, NY, USA},\ \bibinfo {year} {2019})\ p.\ \bibinfo {pages}
  {279–290}\BibitemShut {NoStop}%
\bibitem [{\citenamefont {Arute}\ \emph
  {et~al.}(2019{\natexlab{a}})\citenamefont {Arute}, \citenamefont {Arya},
  \citenamefont {Babbush}, \citenamefont {Bacon}, \citenamefont {Bardin},
  \citenamefont {Barends}, \citenamefont {Biswas}, \citenamefont {Boixo},
  \citenamefont {Brandao}, \citenamefont {Buell} \emph
  {et~al.}}]{arute2019quantum}%
  \BibitemOpen
  \bibfield  {author} {\bibinfo {author} {\bibfnamefont {F.}~\bibnamefont
  {Arute}}, \bibinfo {author} {\bibfnamefont {K.}~\bibnamefont {Arya}},
  \bibinfo {author} {\bibfnamefont {R.}~\bibnamefont {Babbush}}, \bibinfo
  {author} {\bibfnamefont {D.}~\bibnamefont {Bacon}}, \bibinfo {author}
  {\bibfnamefont {J.~C.}\ \bibnamefont {Bardin}}, \bibinfo {author}
  {\bibfnamefont {R.}~\bibnamefont {Barends}}, \bibinfo {author} {\bibfnamefont
  {R.}~\bibnamefont {Biswas}}, \bibinfo {author} {\bibfnamefont
  {S.}~\bibnamefont {Boixo}}, \bibinfo {author} {\bibfnamefont {F.~G.}\
  \bibnamefont {Brandao}}, \bibinfo {author} {\bibfnamefont {D.~A.}\
  \bibnamefont {Buell}}, \emph {et~al.},\ }\bibfield  {title} {\bibinfo {title}
  {Quantum supremacy using a programmable superconducting processor},\
  }\href@noop {} {\bibfield  {journal} {\bibinfo  {journal} {Nature}\ }\textbf
  {\bibinfo {volume} {574}},\ \bibinfo {pages} {505} (\bibinfo {year}
  {2019}{\natexlab{a}})}\BibitemShut {NoStop}%
\bibitem [{\citenamefont {Wu}\ \emph {et~al.}(2021)\citenamefont {Wu},
  \citenamefont {Bao}, \citenamefont {Cao}, \citenamefont {Chen}, \citenamefont
  {Chen}, \citenamefont {Chen}, \citenamefont {Chung}, \citenamefont {Deng},
  \citenamefont {Du}, \citenamefont {Fan} \emph {et~al.}}]{wu2021strong}%
  \BibitemOpen
  \bibfield  {author} {\bibinfo {author} {\bibfnamefont {Y.}~\bibnamefont
  {Wu}}, \bibinfo {author} {\bibfnamefont {W.-S.}\ \bibnamefont {Bao}},
  \bibinfo {author} {\bibfnamefont {S.}~\bibnamefont {Cao}}, \bibinfo {author}
  {\bibfnamefont {F.}~\bibnamefont {Chen}}, \bibinfo {author} {\bibfnamefont
  {M.-C.}\ \bibnamefont {Chen}}, \bibinfo {author} {\bibfnamefont
  {X.}~\bibnamefont {Chen}}, \bibinfo {author} {\bibfnamefont {T.-H.}\
  \bibnamefont {Chung}}, \bibinfo {author} {\bibfnamefont {H.}~\bibnamefont
  {Deng}}, \bibinfo {author} {\bibfnamefont {Y.}~\bibnamefont {Du}}, \bibinfo
  {author} {\bibfnamefont {D.}~\bibnamefont {Fan}}, \emph {et~al.},\ }\bibfield
   {title} {\bibinfo {title} {Strong quantum computational advantage using a
  superconducting quantum processor},\ }\href@noop {} {\bibfield  {journal}
  {\bibinfo  {journal} {arXiv preprint arXiv:2106.14734}\ } (\bibinfo {year}
  {2021})}\BibitemShut {NoStop}%
\bibitem [{\citenamefont {Shor}(1999)}]{shor1999polynomial}%
  \BibitemOpen
  \bibfield  {author} {\bibinfo {author} {\bibfnamefont {P.~W.}\ \bibnamefont
  {Shor}},\ }\bibfield  {title} {\bibinfo {title} {Polynomial-time algorithms
  for prime factorization and discrete logarithms on a quantum computer},\
  }\href@noop {} {\bibfield  {journal} {\bibinfo  {journal} {SIAM review}\
  }\textbf {\bibinfo {volume} {41}},\ \bibinfo {pages} {303} (\bibinfo {year}
  {1999})}\BibitemShut {NoStop}%
\bibitem [{\citenamefont {Nachman}\ \emph {et~al.}(2021)\citenamefont
  {Nachman}, \citenamefont {Provasoli}, \citenamefont {de~Jong},\ and\
  \citenamefont {Bauer}}]{nachman2021quantum}%
  \BibitemOpen
  \bibfield  {author} {\bibinfo {author} {\bibfnamefont {B.}~\bibnamefont
  {Nachman}}, \bibinfo {author} {\bibfnamefont {D.}~\bibnamefont {Provasoli}},
  \bibinfo {author} {\bibfnamefont {W.~A.}\ \bibnamefont {de~Jong}},\ and\
  \bibinfo {author} {\bibfnamefont {C.~W.}\ \bibnamefont {Bauer}},\ }\bibfield
  {title} {\bibinfo {title} {Quantum algorithm for high energy physics
  simulations},\ }\href@noop {} {\bibfield  {journal} {\bibinfo  {journal}
  {Physical Review Letters}\ }\textbf {\bibinfo {volume} {126}},\ \bibinfo
  {pages} {062001} (\bibinfo {year} {2021})}\BibitemShut {NoStop}%
\bibitem [{\citenamefont {Gottesman}(2009)}]{Gottesman09anintroduction}%
  \BibitemOpen
  \bibfield  {author} {\bibinfo {author} {\bibfnamefont {D.}~\bibnamefont
  {Gottesman}},\ }\bibfield  {title} {\bibinfo {title} {An introduction to
  quantum error correction and fault-tolerant quantum computation},\
  }\href@noop {} {\  (\bibinfo {year} {2009})},\ \Eprint
  {https://arxiv.org/abs/0904.2557} {arXiv:0904.2557 [quant-ph]} \BibitemShut
  {NoStop}%
\bibitem [{\citenamefont {Terhal}(2015)}]{terhal2015quantum}%
  \BibitemOpen
  \bibfield  {author} {\bibinfo {author} {\bibfnamefont {B.~M.}\ \bibnamefont
  {Terhal}},\ }\bibfield  {title} {\bibinfo {title} {Quantum error correction
  for quantum memories},\ }\href@noop {} {\bibfield  {journal} {\bibinfo
  {journal} {Reviews of Modern Physics}\ }\textbf {\bibinfo {volume} {87}},\
  \bibinfo {pages} {307} (\bibinfo {year} {2015})}\BibitemShut {NoStop}%
\bibitem [{\citenamefont {Nielsen}\ and\ \citenamefont
  {Chuang}(2011)}]{Nielsen:2011:QCQ:1972505}%
  \BibitemOpen
  \bibfield  {author} {\bibinfo {author} {\bibfnamefont {M.~A.}\ \bibnamefont
  {Nielsen}}\ and\ \bibinfo {author} {\bibfnamefont {I.~L.}\ \bibnamefont
  {Chuang}},\ }\href@noop {} {\emph {\bibinfo {title} {Quantum Computation and
  Quantum Information: 10th Anniversary Edition}}},\ \bibinfo {edition} {10th}\
  ed.\ (\bibinfo  {publisher} {Cambridge University Press},\ \bibinfo {address}
  {New York, NY, USA},\ \bibinfo {year} {2011})\BibitemShut {NoStop}%
\bibitem [{\citenamefont {{M. Urbanek, B. Nachman, and W. de
  Jong}}(2019)}]{errorcorrecting}%
  \BibitemOpen
  \bibfield  {author} {\bibinfo {author} {\bibnamefont {{M. Urbanek, B.
  Nachman, and W. de Jong}}},\ }\bibfield  {title} {\bibinfo {title} {Quantum
  error detection improves accuracy of chemical calculations on a quantum
  computer},\ }\href@noop {} {\  (\bibinfo {year} {2019})},\ \Eprint
  {https://arxiv.org/abs/1910.00129} {arXiv:1910.00129 [quant-ph]} \BibitemShut
  {NoStop}%
\bibitem [{\citenamefont {Wootton}\ and\ \citenamefont
  {Loss}(2018)}]{PhysRevA.97.052313}%
  \BibitemOpen
  \bibfield  {author} {\bibinfo {author} {\bibfnamefont {J.~R.}\ \bibnamefont
  {Wootton}}\ and\ \bibinfo {author} {\bibfnamefont {D.}~\bibnamefont {Loss}},\
  }\bibfield  {title} {\bibinfo {title} {Repetition code of 15 qubits},\ }\href
  {https://doi.org/10.1103/PhysRevA.97.052313} {\bibfield  {journal} {\bibinfo
  {journal} {Phys. Rev. A}\ }\textbf {\bibinfo {volume} {97}},\ \bibinfo
  {pages} {052313} (\bibinfo {year} {2018})}\BibitemShut {NoStop}%
\bibitem [{\citenamefont {Barends}\ \emph {et~al.}(2014)\citenamefont
  {Barends}, \citenamefont {Kelly}, \citenamefont {Megrant}, \citenamefont
  {Veitia}, \citenamefont {Sank}, \citenamefont {Jeffrey}, \citenamefont
  {White}, \citenamefont {Mutus}, \citenamefont {Fowler}, \citenamefont
  {Campbell}, \citenamefont {Chen}, \citenamefont {Chen}, \citenamefont
  {Chiaro}, \citenamefont {Dunsworth}, \citenamefont {Neill}, \citenamefont
  {O'Malley}, \citenamefont {Roushan}, \citenamefont {Vainsencher},
  \citenamefont {Wenner}, \citenamefont {Korotkov}, \citenamefont {Cleland},\
  and\ \citenamefont {Martinis}}]{Barends2014}%
  \BibitemOpen
  \bibfield  {author} {\bibinfo {author} {\bibfnamefont {R.}~\bibnamefont
  {Barends}}, \bibinfo {author} {\bibfnamefont {J.}~\bibnamefont {Kelly}},
  \bibinfo {author} {\bibfnamefont {A.}~\bibnamefont {Megrant}}, \bibinfo
  {author} {\bibfnamefont {A.}~\bibnamefont {Veitia}}, \bibinfo {author}
  {\bibfnamefont {D.}~\bibnamefont {Sank}}, \bibinfo {author} {\bibfnamefont
  {E.}~\bibnamefont {Jeffrey}}, \bibinfo {author} {\bibfnamefont {T.~C.}\
  \bibnamefont {White}}, \bibinfo {author} {\bibfnamefont {J.}~\bibnamefont
  {Mutus}}, \bibinfo {author} {\bibfnamefont {A.~G.}\ \bibnamefont {Fowler}},
  \bibinfo {author} {\bibfnamefont {B.}~\bibnamefont {Campbell}}, \bibinfo
  {author} {\bibfnamefont {Y.}~\bibnamefont {Chen}}, \bibinfo {author}
  {\bibfnamefont {Z.}~\bibnamefont {Chen}}, \bibinfo {author} {\bibfnamefont
  {B.}~\bibnamefont {Chiaro}}, \bibinfo {author} {\bibfnamefont
  {A.}~\bibnamefont {Dunsworth}}, \bibinfo {author} {\bibfnamefont
  {C.}~\bibnamefont {Neill}}, \bibinfo {author} {\bibfnamefont
  {P.}~\bibnamefont {O'Malley}}, \bibinfo {author} {\bibfnamefont
  {P.}~\bibnamefont {Roushan}}, \bibinfo {author} {\bibfnamefont
  {A.}~\bibnamefont {Vainsencher}}, \bibinfo {author} {\bibfnamefont
  {J.}~\bibnamefont {Wenner}}, \bibinfo {author} {\bibfnamefont {A.~N.}\
  \bibnamefont {Korotkov}}, \bibinfo {author} {\bibfnamefont {A.~N.}\
  \bibnamefont {Cleland}},\ and\ \bibinfo {author} {\bibfnamefont {J.~M.}\
  \bibnamefont {Martinis}},\ }\bibfield  {title} {\bibinfo {title}
  {Superconducting quantum circuits at the surface code threshold for fault
  tolerance},\ }\href {https://doi.org/10.1038/nature13171} {\bibfield
  {journal} {\bibinfo  {journal} {Nature}\ }\textbf {\bibinfo {volume} {508}},\
  \bibinfo {pages} {500} (\bibinfo {year} {2014})}\BibitemShut {NoStop}%
\bibitem [{\citenamefont {Kelly}\ \emph {et~al.}(2015)\citenamefont {Kelly},
  \citenamefont {Barends}, \citenamefont {Fowler}, \citenamefont {Megrant},
  \citenamefont {Jeffrey}, \citenamefont {White}, \citenamefont {Sank},
  \citenamefont {Mutus}, \citenamefont {Campbell}, \citenamefont {Chen},
  \citenamefont {Chen}, \citenamefont {Chiaro}, \citenamefont {Dunsworth},
  \citenamefont {Hoi}, \citenamefont {Neill}, \citenamefont {O'Malley},
  \citenamefont {Quintana}, \citenamefont {Roushan}, \citenamefont
  {Vainsencher}, \citenamefont {Wenner}, \citenamefont {Cleland},\ and\
  \citenamefont {Martinis}}]{Kelly2015}%
  \BibitemOpen
  \bibfield  {author} {\bibinfo {author} {\bibfnamefont {J.}~\bibnamefont
  {Kelly}}, \bibinfo {author} {\bibfnamefont {R.}~\bibnamefont {Barends}},
  \bibinfo {author} {\bibfnamefont {A.~G.}\ \bibnamefont {Fowler}}, \bibinfo
  {author} {\bibfnamefont {A.}~\bibnamefont {Megrant}}, \bibinfo {author}
  {\bibfnamefont {E.}~\bibnamefont {Jeffrey}}, \bibinfo {author} {\bibfnamefont
  {T.~C.}\ \bibnamefont {White}}, \bibinfo {author} {\bibfnamefont
  {D.}~\bibnamefont {Sank}}, \bibinfo {author} {\bibfnamefont {J.~Y.}\
  \bibnamefont {Mutus}}, \bibinfo {author} {\bibfnamefont {B.}~\bibnamefont
  {Campbell}}, \bibinfo {author} {\bibfnamefont {Y.}~\bibnamefont {Chen}},
  \bibinfo {author} {\bibfnamefont {Z.}~\bibnamefont {Chen}}, \bibinfo {author}
  {\bibfnamefont {B.}~\bibnamefont {Chiaro}}, \bibinfo {author} {\bibfnamefont
  {A.}~\bibnamefont {Dunsworth}}, \bibinfo {author} {\bibfnamefont {I.-C.}\
  \bibnamefont {Hoi}}, \bibinfo {author} {\bibfnamefont {C.}~\bibnamefont
  {Neill}}, \bibinfo {author} {\bibfnamefont {P.~J.~J.}\ \bibnamefont
  {O'Malley}}, \bibinfo {author} {\bibfnamefont {C.}~\bibnamefont {Quintana}},
  \bibinfo {author} {\bibfnamefont {P.}~\bibnamefont {Roushan}}, \bibinfo
  {author} {\bibfnamefont {A.}~\bibnamefont {Vainsencher}}, \bibinfo {author}
  {\bibfnamefont {J.}~\bibnamefont {Wenner}}, \bibinfo {author} {\bibfnamefont
  {A.~N.}\ \bibnamefont {Cleland}},\ and\ \bibinfo {author} {\bibfnamefont
  {J.~M.}\ \bibnamefont {Martinis}},\ }\bibfield  {title} {\bibinfo {title}
  {State preservation by repetitive error detection in a superconducting
  quantum circuit},\ }\href {https://doi.org/10.1038/nature14270} {\bibfield
  {journal} {\bibinfo  {journal} {Nature}\ }\textbf {\bibinfo {volume} {519}},\
  \bibinfo {pages} {66} (\bibinfo {year} {2015})}\BibitemShut {NoStop}%
\bibitem [{\citenamefont {Linke}\ \emph {et~al.}(2017)\citenamefont {Linke},
  \citenamefont {Gutierrez}, \citenamefont {Landsman}, \citenamefont {Figgatt},
  \citenamefont {Debnath}, \citenamefont {Brown},\ and\ \citenamefont
  {Monroe}}]{Linke:2017}%
  \BibitemOpen
  \bibfield  {author} {\bibinfo {author} {\bibfnamefont {N.~M.}\ \bibnamefont
  {Linke}}, \bibinfo {author} {\bibfnamefont {M.}~\bibnamefont {Gutierrez}},
  \bibinfo {author} {\bibfnamefont {K.~A.}\ \bibnamefont {Landsman}}, \bibinfo
  {author} {\bibfnamefont {C.}~\bibnamefont {Figgatt}}, \bibinfo {author}
  {\bibfnamefont {S.}~\bibnamefont {Debnath}}, \bibinfo {author} {\bibfnamefont
  {K.~R.}\ \bibnamefont {Brown}},\ and\ \bibinfo {author} {\bibfnamefont
  {C.}~\bibnamefont {Monroe}},\ }\bibfield  {title} {\bibinfo {title}
  {Fault-tolerant quantum error detection},\ }\href
  {https://doi.org/10.1126/sciadv.1701074} {\bibfield  {journal} {\bibinfo
  {journal} {Sci. Adv.}\ }\textbf {\bibinfo {volume} {3}},\ \bibinfo {pages}
  {e1701074} (\bibinfo {year} {2017})}\BibitemShut {NoStop}%
\bibitem [{\citenamefont {Takita}\ \emph {et~al.}(2017)\citenamefont {Takita},
  \citenamefont {Cross}, \citenamefont {C\'{o}rcoles}, \citenamefont {Chow},\
  and\ \citenamefont {Gambetta}}]{Takita:2017}%
  \BibitemOpen
  \bibfield  {author} {\bibinfo {author} {\bibfnamefont {M.}~\bibnamefont
  {Takita}}, \bibinfo {author} {\bibfnamefont {A.~W.}\ \bibnamefont {Cross}},
  \bibinfo {author} {\bibfnamefont {A.~D.}\ \bibnamefont {C\'{o}rcoles}},
  \bibinfo {author} {\bibfnamefont {J.~M.}\ \bibnamefont {Chow}},\ and\
  \bibinfo {author} {\bibfnamefont {J.~M.}\ \bibnamefont {Gambetta}},\
  }\bibfield  {title} {\bibinfo {title} {Experimental demonstration of
  fault-tolerant state preparation with superconducting qubits},\ }\href
  {https://doi.org/10.1103/PhysRevLett.119.180501} {\bibfield  {journal}
  {\bibinfo  {journal} {Phys. Rev. Lett.}\ }\textbf {\bibinfo {volume} {119}},\
  \bibinfo {pages} {180501} (\bibinfo {year} {2017})}\BibitemShut {NoStop}%
\bibitem [{\citenamefont {Roffe}\ \emph {et~al.}(2018)\citenamefont {Roffe},
  \citenamefont {Headley}, \citenamefont {Chancellor}, \citenamefont
  {Horsman},\ and\ \citenamefont {Kendon}}]{Roffe:2018}%
  \BibitemOpen
  \bibfield  {author} {\bibinfo {author} {\bibfnamefont {J.}~\bibnamefont
  {Roffe}}, \bibinfo {author} {\bibfnamefont {D.}~\bibnamefont {Headley}},
  \bibinfo {author} {\bibfnamefont {N.}~\bibnamefont {Chancellor}}, \bibinfo
  {author} {\bibfnamefont {D.}~\bibnamefont {Horsman}},\ and\ \bibinfo {author}
  {\bibfnamefont {V.}~\bibnamefont {Kendon}},\ }\bibfield  {title} {\bibinfo
  {title} {Protecting quantum memories using coherent parity check codes},\
  }\href {https://doi.org/10.1088/2058-9565/aac64e} {\bibfield  {journal}
  {\bibinfo  {journal} {Quantum Sci. Technol.}\ }\textbf {\bibinfo {volume}
  {3}},\ \bibinfo {pages} {035010} (\bibinfo {year} {2018})}\BibitemShut
  {NoStop}%
\bibitem [{\citenamefont {Vuillot}(2018)}]{Vuillot:2018}%
  \BibitemOpen
  \bibfield  {author} {\bibinfo {author} {\bibfnamefont {C.}~\bibnamefont
  {Vuillot}},\ }\bibfield  {title} {\bibinfo {title} {Is error detection
  helpful on {IBM 5Q} chips?},\ }\href {https://doi.org/10.26421/qic18.11-12}
  {\bibfield  {journal} {\bibinfo  {journal} {Quantum Inf. Comput.}\ }\textbf
  {\bibinfo {volume} {18}},\ \bibinfo {pages} {0949} (\bibinfo {year}
  {2018})}\BibitemShut {NoStop}%
\bibitem [{\citenamefont {Willsch}\ \emph {et~al.}(2018)\citenamefont
  {Willsch}, \citenamefont {Willsch}, \citenamefont {Jin}, \citenamefont
  {De~Raedt},\ and\ \citenamefont {Michielsen}}]{Willsch:2018}%
  \BibitemOpen
  \bibfield  {author} {\bibinfo {author} {\bibfnamefont {D.}~\bibnamefont
  {Willsch}}, \bibinfo {author} {\bibfnamefont {M.}~\bibnamefont {Willsch}},
  \bibinfo {author} {\bibfnamefont {F.}~\bibnamefont {Jin}}, \bibinfo {author}
  {\bibfnamefont {H.}~\bibnamefont {De~Raedt}},\ and\ \bibinfo {author}
  {\bibfnamefont {K.}~\bibnamefont {Michielsen}},\ }\bibfield  {title}
  {\bibinfo {title} {Testing quantum fault tolerance on small systems},\ }\href
  {https://doi.org/10.1103/PhysRevA.98.052348} {\bibfield  {journal} {\bibinfo
  {journal} {Phys. Rev. A}\ }\textbf {\bibinfo {volume} {98}},\ \bibinfo
  {pages} {052348} (\bibinfo {year} {2018})}\BibitemShut {NoStop}%
\bibitem [{\citenamefont {Harper}\ and\ \citenamefont
  {Flammia}(2019)}]{Harper:2019}%
  \BibitemOpen
  \bibfield  {author} {\bibinfo {author} {\bibfnamefont {R.}~\bibnamefont
  {Harper}}\ and\ \bibinfo {author} {\bibfnamefont {S.~T.}\ \bibnamefont
  {Flammia}},\ }\bibfield  {title} {\bibinfo {title} {Fault-tolerant logical
  gates in the {IBM Quantum Experience}},\ }\href
  {https://doi.org/10.1103/PhysRevLett.122.080504} {\bibfield  {journal}
  {\bibinfo  {journal} {Phys. Rev. Lett.}\ }\textbf {\bibinfo {volume} {122}},\
  \bibinfo {pages} {080504} (\bibinfo {year} {2019})}\BibitemShut {NoStop}%
\bibitem [{\citenamefont {Chen}\ \emph {et~al.}(2021)\citenamefont {Chen} \emph
  {et~al.}}]{chen2021exponential}%
  \BibitemOpen
  \bibfield  {author} {\bibinfo {author} {\bibfnamefont {Z.}~\bibnamefont
  {Chen}} \emph {et~al.},\ }\href@noop {} {\bibinfo {title} {Exponential
  suppression of bit or phase flip errors with repetitive error correction}}
  (\bibinfo {year} {2021}),\ \Eprint {https://arxiv.org/abs/2102.06132}
  {arXiv:2102.06132 [quant-ph]} \BibitemShut {NoStop}%
\bibitem [{\citenamefont {Campbell}\ \emph {et~al.}(2017)\citenamefont
  {Campbell}, \citenamefont {Terhal},\ and\ \citenamefont
  {Vuillot}}]{campbell2017roads}%
  \BibitemOpen
  \bibfield  {author} {\bibinfo {author} {\bibfnamefont {E.~T.}\ \bibnamefont
  {Campbell}}, \bibinfo {author} {\bibfnamefont {B.~M.}\ \bibnamefont
  {Terhal}},\ and\ \bibinfo {author} {\bibfnamefont {C.}~\bibnamefont
  {Vuillot}},\ }\bibfield  {title} {\bibinfo {title} {Roads towards
  fault-tolerant universal quantum computation},\ }\href@noop {} {\bibfield
  {journal} {\bibinfo  {journal} {Nature}\ }\textbf {\bibinfo {volume} {549}},\
  \bibinfo {pages} {172} (\bibinfo {year} {2017})}\BibitemShut {NoStop}%
\bibitem [{\citenamefont {Gidney}\ and\ \citenamefont
  {Eker{\aa}}(2021)}]{gidney2021factor}%
  \BibitemOpen
  \bibfield  {author} {\bibinfo {author} {\bibfnamefont {C.}~\bibnamefont
  {Gidney}}\ and\ \bibinfo {author} {\bibfnamefont {M.}~\bibnamefont
  {Eker{\aa}}},\ }\bibfield  {title} {\bibinfo {title} {How to factor 2048 bit
  rsa integers in 8 hours using 20 million noisy qubits},\ }\href@noop {}
  {\bibfield  {journal} {\bibinfo  {journal} {Quantum}\ }\textbf {\bibinfo
  {volume} {5}},\ \bibinfo {pages} {433} (\bibinfo {year} {2021})}\BibitemShut
  {NoStop}%
\bibitem [{\citenamefont {Arute}\ \emph
  {et~al.}(2019{\natexlab{b}})\citenamefont {Arute}, \citenamefont {Arya},
  \citenamefont {Babbush} \emph {et~al.}}]{Sycamore}%
  \BibitemOpen
  \bibfield  {author} {\bibinfo {author} {\bibfnamefont {F.}~\bibnamefont
  {Arute}}, \bibinfo {author} {\bibfnamefont {K.}~\bibnamefont {Arya}},
  \bibinfo {author} {\bibfnamefont {R.}~\bibnamefont {Babbush}}, \emph
  {et~al.},\ }\bibfield  {title} {\bibinfo {title} {Quantum supremacy using a
  programmable superconducting processor},\ }\href
  {https://doi.org/10.1038/s41586-019-1666-5} {\bibfield  {journal} {\bibinfo
  {journal} {Nature}\ }\textbf {\bibinfo {volume} {574}},\ \bibinfo {pages}
  {505–510} (\bibinfo {year} {2019}{\natexlab{b}})}\BibitemShut {NoStop}%
\bibitem [{err()}]{error_model}%
  \BibitemOpen
  \href@noop {} {}\bibinfo {note} {In particular, other two-qubit error models
  would simply modify the linear prefactor $\alpha$ (e.g.~see Table
  \ref{tab:Tab1}), which is determined by the relative likelihood of bit-flip
  errors occurring on \emph{both} qubits. Furthermore, additional single qubit
  errors occuring \emph{after} the \CNOT~gate would not contribute to
  $q_\textrm{eff}$ at leading order, as two independent errors would be
  necessary to change the logical state.}\BibitemShut {Stop}%
\bibitem [{asy()}]{asymm_noise}%
  \BibitemOpen
  \href@noop {} {}\bibinfo {note} {We note that in the highly asymmetric case,
  $\kappa \approx 1$, it may be advantageous to \emph{correct} errors by
  assigning $01$ and $10$ to the logical $\hat 1$ state, instead of discarding
  them. This reduces the measurement overhead associated with post-selection;
  however, it does not remove the $\mathcal{O}(\epsilon)$ error from two qubit
  flips in the encoding circuit.}\BibitemShut {Stop}%
\bibitem [{\citenamefont {Hicks}\ \emph {et~al.}(2021)\citenamefont {Hicks},
  \citenamefont {Bauer},\ and\ \citenamefont {Nachman}}]{Hicks_2021}%
  \BibitemOpen
  \bibfield  {author} {\bibinfo {author} {\bibfnamefont {R.}~\bibnamefont
  {Hicks}}, \bibinfo {author} {\bibfnamefont {C.~W.}\ \bibnamefont {Bauer}},\
  and\ \bibinfo {author} {\bibfnamefont {B.}~\bibnamefont {Nachman}},\
  }\bibfield  {title} {\bibinfo {title} {Readout rebalancing for near-term
  quantum computers},\ }\bibfield  {journal} {\bibinfo  {journal} {Physical
  Review A}\ }\textbf {\bibinfo {volume} {103}},\ \href
  {https://doi.org/10.1103/physreva.103.022407} {10.1103/physreva.103.022407}
  (\bibinfo {year} {2021})\BibitemShut {NoStop}%
\bibitem [{cir()}]{cirq}%
  \BibitemOpen
  \bibfield  {title} {\bibinfo {title} {Cirq, a python framework for creating,
  editing, and invoking noisy intermediate scale quantum (nisq) circuits},\
  }\href@noop {} {\ }\bibinfo {note}
  {\url{https://github.com/quantumlib/Cirq}}\BibitemShut {NoStop}%
\bibitem [{\citenamefont {G{\"u}nther}\ \emph {et~al.}(2021)\citenamefont
  {G{\"u}nther}, \citenamefont {Tacchino}, \citenamefont {Wootton},
  \citenamefont {Tavernelli},\ and\ \citenamefont
  {Barkoutsos}}]{gunther2021improving}%
  \BibitemOpen
  \bibfield  {author} {\bibinfo {author} {\bibfnamefont {J.~M.}\ \bibnamefont
  {G{\"u}nther}}, \bibinfo {author} {\bibfnamefont {F.}~\bibnamefont
  {Tacchino}}, \bibinfo {author} {\bibfnamefont {J.~R.}\ \bibnamefont
  {Wootton}}, \bibinfo {author} {\bibfnamefont {I.}~\bibnamefont
  {Tavernelli}},\ and\ \bibinfo {author} {\bibfnamefont {P.~K.}\ \bibnamefont
  {Barkoutsos}},\ }\bibfield  {title} {\bibinfo {title} {Improving readout in
  quantum simulations with repetition codes},\ }\href@noop {} {\bibfield
  {journal} {\bibinfo  {journal} {arXiv preprint arXiv:2105.13377}\ } (\bibinfo
  {year} {2021})}\BibitemShut {NoStop}%
\end{thebibliography}%

\clearpage
\appendix

\section{Readout variance with error mitigation}\label{app:var}

While we have mostly focused on the bias of quantum observables in the presence of readout errors, it is also important to consider the change in variance under active error mitigation, as this determines the number of measurements required to estimate an observable to a desired degree of precision. 
To this end, let us revisit the single qubit example presented in Section \ref{sec:two-qubit}.
We recall that, without any error mitigation, the distribution of measurements characterized by the expected frequencies $\mathbb{E}[m_0] = 1- \mathbb{E}[m_1] = \lambda$ for $\lambda=p+q(1-2p)$ and the variance $\textrm{Var}[m_0] = \textrm{Var}[m_1] = \lambda(1-\lambda)$.
In practice, we care about the estimator of the frequencies after $N$ measurements, e.g.~$\theta_0 = N_0 / N$ where $N_0$ is the number of measurements in the 0 state.
The variance of the estimator is given by
\begin{align}
\begin{split} \label{var_theta0}
  \textrm{Var}[\theta_0] &= \textrm{Var}[m_0]/N \\
  &= \frac 1 N \lambda(1-\lambda) \\
  &= \frac 1 N \left [p(1-p)+q(1-2p)^2\right] + \mathcal{O}(q^2),
 \end{split}
\end{align}
which increases as a function of the error rate. 

Let us now compare to case of active readout error \emph{detection}.
As we found in the main text, the bias in the distribution of the two logical states, $\bar 0$ and $\bar 1$, is renormalized by the effective error rate: $\mathbb{E}[m_{\bar 0}] = 1- \mathbb{E}[m_{\bar 1}] = \lambda_{\textrm{eff}}$ with $\lambda_\textrm{eff} = p+q_\textrm{eff}(1-2p) + \mathcal{O}(q^2)$.
However, the estimator of these frequencies depends on two competing factors: the change in bias and the total number of measurements that are kept during post-selection.
Specifically, we have $\theta_{\bar 0} = N_{00} / (N_{00}+N_{11}) = N_{00}/(N(1-2q))$.
This leads to a variance
\begin{align}
\begin{split} \label{var_theta00}
  \textrm{Var}[\theta_{\bar 0}] &\approx \frac 1 N \lambda_\textrm{eff}(1-\lambda_\textrm{eff})(1+2q) \\ 
  &\approx \frac 1 N \left [p(1-p)+q_\textrm{eff}(1-2p)^2 + 2qp(1-p) \right],
 \end{split}
\end{align}
where we have dropped $\mathcal{O}(q^2,q_\textrm{eff}^2)$ terms.
Comparing with Eq.~\eqref{var_theta0}, we find that whether or not error detection improves the estimator variance depends on both the state parameter $p$ and the ratio $q_\textrm{eff} / q$.

In contrast, when error \emph{correction} is performed, no measurements are discarded.
Hence, the variance of the estimator for error correction depends only on $q_\textrm{eff}$:
\begin{align}
\begin{split} \label{var_cor}
  \textrm{Var}[\theta_{\bar 0}] &= \frac 1 N \lambda_\textrm{eff}(1-\lambda_\textrm{eff}) \\ 
  &= \frac 1 N \left [p(1-p)+q_\textrm{eff}(1-2p)^2\right]  + \mathcal{O}(q_\textrm{eff}^2).
 \end{split}
\end{align}
This is strictly lower than unmitigated variance [Eq.~\eqref{var_theta0}] whenever $q_\textrm{eff} < q$ and, for certain parameters, can be lower than variance with error detection [Eq.~\eqref{var_theta00}].

Finally, we note that the change in variance with or without error mitigation is generally small compared to intrinsic quantum uncertainty, i.e.~$p(1-p)$.
For this reason, the more significant benefits offered by active error mitigation may be its application to problems where passive readout mitigation methods are unsuitable, e.g.~when a response matrix cannot be accurately determined or the output of individual shots is required.

\section{Review of Hamming codes}\label{app:Hamming}

In this appendix, we briefly describe the family of classical error correction codes known as Hamming codes. 
The 7-qubit code considered in this work is a member of this family; in fact, the three-qubit repetition code is a member as well (though, for clarity, we have chosen not to use this terminology).
%
%
Each Hamming code has a code distance of $d=3$, implying that a single bit-flip error can be corrected and up to two bit-flip errors can be detected.
%
%
Specifically, for any integer $r \ge 2$, a Hamming code can be constructed to encode $2^r-r-1$ logical bits into $2^r - 1$ physical bits ($r=2$ is the 3-qubit repetition code; $r=3$ is the 7-qubit code).
Thus, as $r$ increases, the ratio of physical to logical bits quickly approaches one.

The general procedure for constructing a Hamming code is as follows.
One begins by enumerating the $2^r - 1$ physical bits starting from $l =  1,2,...,2^r - 1$ and assigning the bits with index $2^i$ for $i = 0, 1,\ldots,r-1$ as parity bits and the rest as logical bits.
The role of each parity bit is to  store the parity of a particular set of logical bits. In particular, for the $i$th parity bit, this set includes all logical bits whose index $l$ as a binary string satisfies $l_i = 1$ (i.e.~the $i$th bit of $l$ is equal to 1).  
For example, in the (7,4) code, the $l=1$ parity bit stores the parity of the logical bits with index 3, 5, and 7.
 
%
%
To utilize the encoding, one checks if the state of each parity bit matches the measured parity of the corresponding set of logical bits; this is accomplished by multiplying the measurement outcome by the parity check matrix, as shown in Eq.~\eqref{eq:parity check} for the (7,4) code.
If only a single bit-flip error has occurred, the index of the error is given by the binary string whose bits are set to 1 for each parity check that is violated.
For example, if the $l=1$ and $l=2$ parity checks are violated, it indicates that an error occurred on the logical bit with index $l=3$; while if only the $l=1$ parity check is violated, it indicates that this parity bit itself had an error.
However, if more than one bit-flip errors occur, the errors can no longer be correctly identified (though they can still be detected for up to two bit-flip errors); indeed, attempting to perform error correction would lead to a spurious final state.

Finally, we note that, for each Hamming code, one can form an extended Hamming code by adding a single extra parity bit, denoted with the label $l=0$, which stores the parity of \emph{all} the physical bits..
The effect of this extra parity bit is to increase the code distance from $d=3$ to $d=4$.
We provide a detailed analysis of the Hamming (8,4) code in Appendix \ref{app:8-hamming}.

\section{Analysis of gate errors}\label{app:gate}

A key feature of active readout error correction is that the effective readout error rate remains linearly susceptible to gate errors during the encoding circuit, i.e.~$Q_{\textrm{eff}} / k \approx \alpha \epsilon$, when $\epsilon, q \ll 1$.
This is because single gate errors can lead to correlated bit-flip errors in the final measured outcomes.
%
As we discussed in Section~\ref{sec:results}, the prefactor $\alpha$ can be derived by summing all error channels that lead to a logical error, weighted by their individual likelihood (e.g.~$\epsilon/16$ for a two-qubit symmetric depolarizing channel).
Note that the definition of a logical error depends both on the encoding and also whether one is performing error detection or correction, as in the latter case it suffices to falsely \emph{identify} an error.

%

%
\begin{figure}
    \centering
    \includegraphics[width=0.3\textwidth]{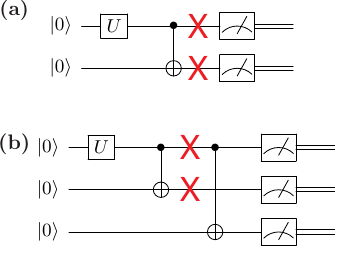}
    \caption{Schematic of the single undetectable error channel for (a) the (2,1) code, and (b) the (3,1) code. The red crosses indicate the location of a bit-flip error (i.e.~a Pauli $X$- or $Y$-type error).}
   \label{fig:rem_error_det}
\end{figure}

\begin{figure}
    \centering
    \includegraphics[width=0.3\textwidth]{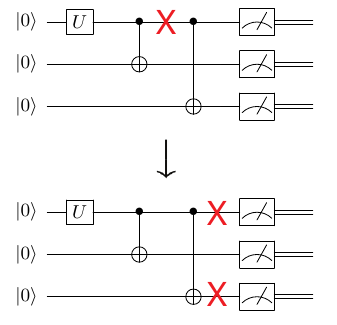}
    \caption{Example of an uncorrectable error for the (3,1) code. The error occurs on a single site (top), but propagates to a second site (bottom) due to the second \CNOT~gate.}
    \label{fig:rem_error_cor}
\end{figure}

\begin{figure}
    \centering
    \includegraphics[width=0.35\textwidth]{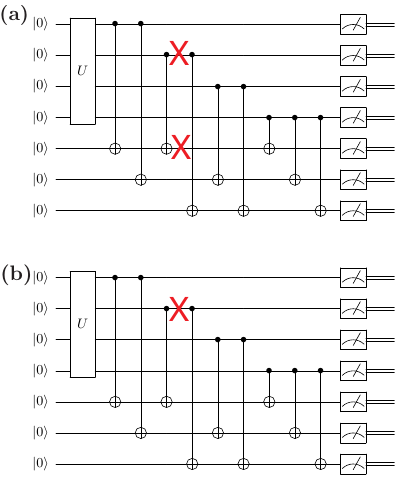}
    \caption{Example of (a) an undetectable and (b) an uncorrectable error for the (7,4) code.}
    \label{fig:hamm_error}
\end{figure}

We previously identified the relevant error channels for the (2,1) and (3,1) codes.
For the (2,1) code, they consist of correlated errors that flip both of the physical bits, i.e.~Pauli errors of the form XX, XY, YX, or YY [Fig.~\ref{fig:rem_error_det}].
The (3,1) code with error detection is directly analagous: logical errors require all of the physical bits to be flipped and are caused only by correlated errors following the first \CNOT~gate.
However, for the (3,1) code with error correction, logical errors require only two out of the three qubits to be flipped; this can occur by a two-qubit error on the second \CNOT~gate or single qubit error after the first \CNOT~gate (i.e.~$X$ or $Y$ error) which propagates into a two-qubit error, as depicted in Fig.~\ref{fig:rem_error_cor}.
Adding together these channels, we determine the values for $\alpha$ presented in Table \ref{tab:Tab1} and confirmed numerically in Fig.~\ref{fig:err_scaling}.

We now extend this logic to analyze the Hamming (7,4) code. 
In order to have a fully \emph{undetectable} logical error, all parity checks must remain unviolated, requiring that a bit-flip error occurs on one of the logical bits and \emph{all} of the parity bits associated with it.
As with repetition codes, this situation can only arise through a correlated two-qubit error after the first \CNOT~gate acting on one of the logical bits [Fig.~\ref{fig:hamm_error}(a)]. 
Indeed, such an error is equivalent to having a bit flip on the logical qubit itself \emph{before} the encoding circuit, making it clearly undetectable. 
On the other hand, having a falsely \emph{identified} logical error, requires only that a spurious combination of parity checks to be violated.
This can occur in two possible ways.
First, a two-qubit bit-flip error can occur after \emph{any} of the \CNOT~gates (including the undetectable error mentioned above).
Second, a single-qubit bit-flip error can occur after one \CNOT~gate and propogate into a multi-qubit error owing to a later \CNOT~gate [Fig.~\ref{fig:hamm_error}(b)].
For this, the single-qubit error must occur on the control bit of a subsequent \CNOT~gate.

This understanding allows us to directly count the number of relevant error channels for the (7,4) code.
Referring to Fig.~\ref{fig:schematic3}, there are a total of 9 \CNOT~gates which could directly lead to a two-qubit error, while there are 4 gates for which a single qubit error could later propogate.
We mutiply these by the number of distinct Pauli errors per gate (4 two-qubit errors and 2 single-qubit errors), and weight them by their individual likelihood of occuring.
This yields a total error rate of $Q_{\textrm{eff}} = \frac{\epsilon} {16} (4*9+2*4) = 4*\frac 7 2$, which agrees precisely with the numerical results presented in Fig.~\ref{fig:err_scaling}.
 
We can further generalize this counting argument to predict the total error rate for an arbitrary Hamming code, denoted a $(2^r-1,2^r-r-1)$ code . 
To do so, we note that the layout \CNOT~gates is given by the binary representation of each logical bit.
Thus, the number of \CNOT~gates connected to a logical bit is equal to the number of 1s in the binary string, and the number of gates that could lead to a propagating error is equal to one fewer (i.e.~only excludes the last gate).
Summing each of the logical bits, the total number of \CNOT~gates is
\begin{align*}
\sum_{m=2}^r {r \choose m}k &= r 2^{r-1} - r \\
& = (r-1)2^{r-1}
\end{align*}
and the total number of propagating errors is
\begin{align*}
\sum_{m=2}^r {r \choose m}(m-1) &= r 2^{r-1} - r - (2^r - r). \\
&= (r-1)2^{r-2}
\end{align*}
This yields a total error rate
\begin{equation*}
Q_{\textrm{eff}} = \frac \epsilon {16}\left [4 (r-1)2^{r-1} + 2 (r-1)2^{r-1} \right],
\end{equation*}
which, for $r \gg 1$, we can approximate as 
\begin{equation*}
Q_{\textrm{eff}} \approx \frac{\epsilon} {16} \left(4r2^{r-1}+2r2^{r-1} \right).
\end{equation*}
Recalling that the number of logical qubits is $k\approx 2^r$, we can express the error rate as $Q_{\textrm{eff}} /k \approx \alpha \epsilon$ with $\alpha \approx 3r/16 \approx 3/16 \log_2 n$.
Incidentally, we could also have arrived at this result by realizing that the number of \CNOT~gates per logical qubit is sharply peaked at $r/2$, corresponding to the typical number of 1's in a random binary string.
 
We conclude that, when performing error correction, the error rate per qubit increases logarithmically with the total number of qubits.
This is in contrast to the \emph{constant} error rate per qubit, i.e.~$\alpha = 1/4$, when performing error detection.
While the exact error rates depended on our model of 2-qubit depolarizing noise, we expect these trends to hold for arbitrary noise models.

\section{Hamming (8,4) code}\label{app:8-hamming}

\begin{figure}
    \centering
    \includegraphics[width=0.48\textwidth]{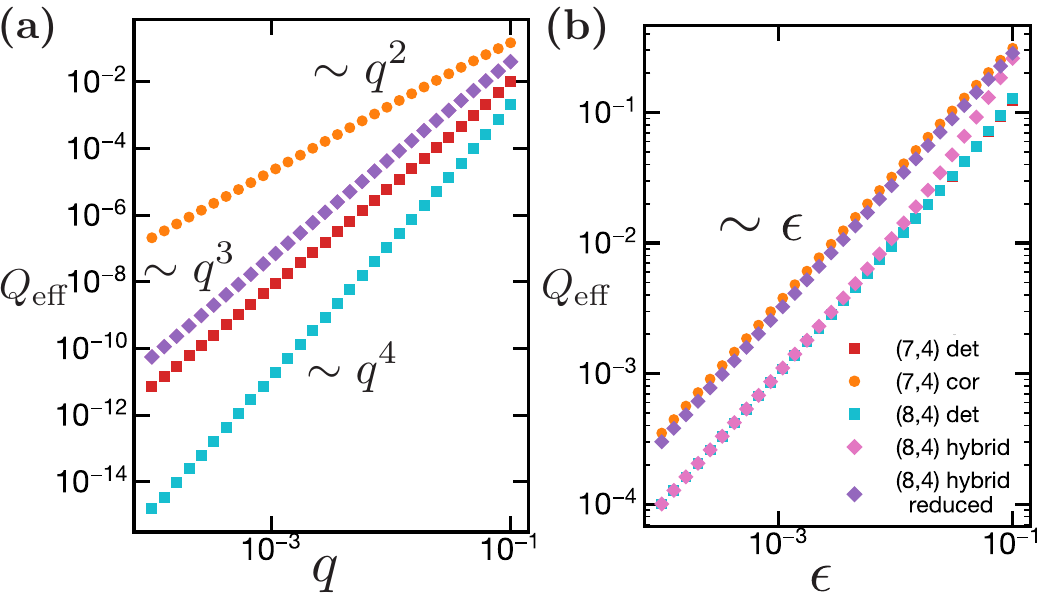}
    \caption{Total readout error rate, $Q_{\textrm{eff}}$, for different error mitigation strategies, analogous to Fig.~\ref{fig:err_scaling}. Note the ``hybrid'' label for the (8,4) code indicates that single bit-flip errors are corrected, while others errors are detected and discarded. The ``reduced'' circuit for the (8,4) code is shown in Fig.~\ref{fig:8hamming_reduced}.}
    \label{fig:err_scaling2}
\end{figure}

As mentioned in Appendix \ref{app:Hamming}, an extended version of each Hamming code may be obtained by including an additional parity bit that stores the parity of all logical bits.
This increases the code distance from $d=3$ to $d=4$, enabling the detection of up to three bit-flip errors and the correction of only single bit-flip, as in the non-extended version.
Notably, this implies that two-qubit errors can be \emph{distinguished} from single-qubit errors; in contrast, in the non-extended version, two-qubit errors can always be misidentified as a single-qubit error.
Thus, a common usage for the code is to perform a hybrid between error correction and error detection, i.e.~one corrects ostensibly single-qubit errors and discards multi-qubit errors.

We consider the implementation of this code using the encoding circuit shown in Fig.~\ref{fig:schematic3}.
The corresponding parity check matrix is given by:
\begin{equation}
H = 
\begin{pmatrix}
1 & 1 & 1 & 1 & 1 & 1 & 1 & 1 \\
1 & 1 & 0 & 1 & 1 & 0 & 0 & 0 \\
1 & 0 & 1 & 1 & 0 & 1 & 0 & 0 \\
0 & 1 & 1 & 1 & 0 & 0 & 1 & 0 \\
\end{pmatrix}_{[8,4]} .
\end{equation}
If the first parity check is violated, it suggests that a single bit-flip error has occurred, and the remaining parity checks can locate the error in the same way as the non-extended version (of course, the same scenario could occur with three bit-flip errors).
On the other hand, if the first parity check succeeds, but one of the other checks fail, it indicates that multiple bit-flip errors have occurred, but because the location of these cannot be determined, the measurement is discarded.

In Fig.~\ref{fig:err_scaling2}, we test the performance of this hybrid approach via noisy simulations with the same noise channels as in main text. 
We first verify that the total error rate scales as $Q_{\textrm{eff}} \sim q^3$ with respect to readout noise, as expected from the fact that three bit-flip errors can be misidentified as a single bit-flip error.
More surprisingly, the linear susceptibility with respect to gate errors is $Q_{\textrm{eff}} /k = \alpha \epsilon$ with $\alpha = 1/4$, the same as other codes performing pure error detection.
While this suggest that the hybrid approach is more robust to gate errors, a close inspection of the circuit reveals that all gate errors that plagued the (7,4) code with error correction are not \emph{correctable} by the (8,4) code; rather they contribute to the number of discarded measurements.
More specifically, such errors propagate during the last round of \CNOT~gates (shown in the box of Fig.~\ref{fig:schematic3}), resulting in least two bit-flip errors on the measurement outcome, which cannot be corrected. 
Indeed, we verify that the number of discarded measurements for the (8,4) code is equivalent to that of the (7,4) code with error detection when only gate errors are included.
Thus, utilizing the (8,4) code with the hybrid error correction / detection strategy is practically equivalent to utilizing the (7,4) code with error detection in terms of handling gate errors (as well as readout errors).

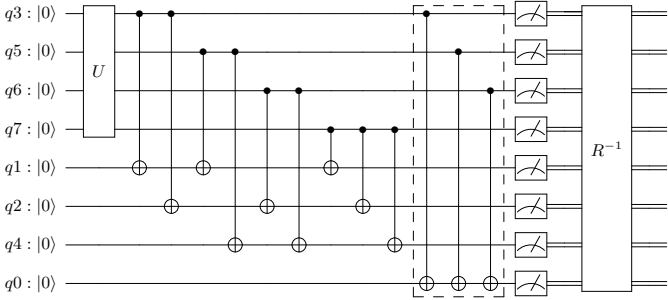
\begin{figure}[b]
\adjustbox{max width=0.45\textwidth}{
\Qcircuit @C=1em @R=0.8em {
    \lstick{q3:\ket{0}}  & \multigate{3}{U}  &\ctrl{4}&\ctrl{5}&\qw&\qw&\qw&\qw&\qw&\qw&\qw& \ctrl{7}&\qw&\qw&\meter & \cw \cw[1] &\cmultigate{7}{R^{-1}} &\cw &\cw \\
    \lstick{q5:\ket{0}}  & \ghost{U} & \qw & \qw&\ctrl{3}&\ctrl{5}&\qw&\qw&\qw&\qw&\qw&\qw&\ctrl{6} & \qw& \meter & \cw \cw[1] &\pureghost{R^{-1}} &\cw &\cw \\
    \lstick{q6:\ket{0}}  & \ghost{U}  & \qw& \qw &\qw&\qw&\ctrl{3}&\ctrl{4}&\qw&\qw&\qw&\qw&\qw&\ctrl{5}& \meter & \cw \cw[1] &\pureghost{R^{-1}} &\cw &\cw \\
    \lstick{q7:\ket{0}}  & \ghost{U} & \qw& \qw &\qw&\qw&\qw&\qw&\ctrl{1}&\ctrl{2}&\ctrl{3}& \qw&\qw&\qw&\meter & \cw \cw[1] &\pureghost{R^{-1}} &\cw &\cw \\
    \lstick{q1:\ket{0}}  & \qw& \targ & \qw&\targ& \qw&\qw&\qw&\targ&\qw&\qw&\qw&\qw&\qw&\meter & \cw \cw[1] &\pureghost{R^{-1}} &\cw &\cw \\
    \lstick{q2:\ket{0}}  & \qw & \qw& \targ &\qw& \qw&\targ&\qw&\qw&\targ&\qw&\qw&\qw&\qw &\meter & \cw \cw[1] &\pureghost{R^{-1}} &\cw &\cw \\
    \lstick{q4:\ket{0}}  & \qw & \qw& \qw &\qw& \targ&\qw &\targ&\qw&\qw&\targ&\qw&\qw&\qw&\meter & \cw \cw[1] &\pureghost{R^{-1}} &\cw &\cw \\
     \lstick{q0:\ket{0}}  & \qw & \qw& \qw &\qw& \qw&\qw &\qw&\qw&\qw&\qw&\targ&\targ&\targ&\meter & \cw \cw[1] &\pureghost{R^{-1}} &\cw &\cw \gategroup{1}{12}{8}{14}{.7em}{--} \\
      }
}
\caption{Reduced circuit for encoding the Hamming (8,4) code.}
\label{fig:8hamming_reduced}
\end{figure}

Finally, we note that a simpler circuit can, in fact, be utilized to implement the (8,4) code.
The reason is that the parity bits redundantly stores the parity of certain combinations of logical bits, so one can obtain the parity of all the physical bits by interacting with only a partial set of the bits. 
The reduced circuit is shown in Fig.~\ref{fig:8hamming_reduced} and consists of only 3 \CNOT~gates acting on the last parity bit rather than the original 7 gates.
By performing noisy simulations, we verify that this reduced circuit has the same scaling with respect to readout errors as the full circuit.
However, the simplified circuit exhibits a \emph{worse} performance with respect to gate errors; in particular, we find $\alpha = 3/4$.
The explanation for this counter-intuitive result is that, in the simplified circuit, many of gate errors that occur in the non-extended part of the circuit do not propagate during the final round of \CNOT~gates. 
As a result, many such errors are misidentified as single bit-flip errors and lead to a spurious final state, similar to the (7,4) code performing error correction. 
In contrast, for the full circuit shown in Fig.~\ref{fig:schematic3}, these same errors always propagate and are flagged as multi-qubit errors which are discarded.

\appendix

\end{document}